\documentclass[11pt,a4paper]{article}

\usepackage[utf8]{inputenc}
\usepackage[margin=1.25in]{geometry}
\usepackage[T1]{fontenc}
\usepackage[english]{babel}
\usepackage{color}
\usepackage[textsize=small]{todonotes}
\usepackage[algo2e,ruled,lined,linesnumberedhidden,longend]{algorithm2e}
    \SetArgSty{textsl}
    \SetCommentSty{textnormal}
    \SetKwFor{For}{for}{}{endfor}
\usepackage[cmex10]{amsmath}
\usepackage{amssymb}
\usepackage{amsfonts}
\usepackage{amsthm}
\usepackage{enumerate}
\usepackage{cite}
\usepackage{multicol}
\usepackage{multirow}
\usepackage{fullpage}
\usepackage{hyperref}

\usepackage[all,dvips]{xy}
\usepackage{graphicx}
\usepackage{tikz}
\usetikzlibrary{matrix,arrows,decorations.pathmorphing}

\theoremstyle{plain}
\newtheorem{theorem}{Theorem}
\newtheorem{proposition}[theorem]{Proposition}
\newtheorem{corollary}[theorem]{Corollary}
\newtheorem{lemma}[theorem]{Lemma}
\newtheorem{conjecture}{Conjecture}

\theoremstyle{definition}
\newtheorem{definition}{Definition}
\newtheorem{notation}{Notation}

\theoremstyle{remark}
\newtheorem{remark}{Remark}
\newtheorem{example}{Example}

\usepackage{algorithmic}

\newcommand{\C}[3]{\mathcal C_L(\mathcal{#1},\mathcal {#2},{#3})}
\newcommand{\CC}[3]{\mathcal C_L(#1,#2,#3)}
\newcommand{\X}{\mathcal{X}}
\newcommand{\F}{\mathbb{F}}
\newcommand{\Fq}{\mathbb{F}_q}
\newcommand{\Z}{\mathbb{Z}}

\renewcommand{\P}{\mathcal{P}}
\newcommand{\DP}{D_{\mathcal{P}}}
\newcommand{\DPP}{D_{\mathcal{P'}}}

\newcommand{\ab}{\mathbf{a}}
\newcommand{\bb}{\mathbf{b}}
\newcommand{\cb}{\mathbf{c}}

\newcommand{\Gm}{\mathbf{G}}

\renewcommand{\leq}{\leqslant}
\renewcommand{\geq}{\geqslant}
\newcommand{\map}[4]{\left\{
    \begin{array}{ccc}
      #1 & \rightarrow & #2\\ #3 & \mapsto & #4
    \end{array}
\right.}
\newcommand{\eqdef }{:=}
\newcommand{\ev}{\textrm{ev}}
\newcommand{\Span}{\textrm{Span}}
\usepackage{authblk}
\newcommand*\samethanks[1][\value{footnote}]{\footnotemark[#1]}

\begin{document}

\title{Cryptanalysis of McEliece Cryptosystem Based on Algebraic Geometry Codes and their Subcodes}

\author[1]{Alain Couvreur\thanks{Partially funded by ANR grant
    ANR-15-CE39-0013-01 ``Manta'' and by the Commission of the
    European community the Horizon 2020 program ICT--645622
    ``PQCRYPTO''.}} \author[2]{Irene
  M\'arquez-Corbella\samethanks\thanks{Partially funded by Spanish grants
    MTM2016-80659-P and MTM2016-80659-P}} \author[3]{Ruud Pellikaan}
\affil[1]{INRIA, Laboratoire LIX, École Polytechnique \& CNRS UMR
  7161, Université Paris--Saclay.  \'Ecole Polytechnique,
  91128 Palaiseau Cedex - France\\
  alain.couvreur@lix.polytechnique.fr}

\affil[2]{University of La Laguna, 
Dept. Maths., Stats. and O.R.,  
38271 La Laguna - Spain\\
imarquec@ull.es}

\affil[3]{Eindhoven University of Technology
P.O. Box 513,
5600 MB Eindhoven\\
g.r.pellikaan@tue.nl}

\maketitle

 % {\tiny
 % \tableofcontents
 % }

\begin{abstract}
  We give polynomial time attacks on the McEliece public key
  cryptosystem based either on algebraic geometry (AG) codes or on
  small codimensional subcodes of AG codes.  These attacks consist in
  the blind reconstruction either of an \emph{Error Correcting Pair
    (ECP)}, or an \emph{Error Correcting Array (ECA)} from the single
  data of an arbitrary generator matrix of a code.  An ECP provides a
  decoding algorithm that corrects up to $\frac{d^*-1-g}{2}$ errors,
  where $d^*$ denotes the designed distance and $g$ denotes the genus
  of the corresponding curve, while with an ECA the decoding algorithm
  corrects up to $\frac{d^*-1}{2}$ errors.  Roughly speaking, for a
  public code of length $n$ over $\F_q$, these attacks run in
  $O(n^4\log (n))$ operations in $\mathbb F_q$ for the reconstruction
  of an ECP and $O(n^5)$ operations for the reconstruction of an
  ECA. A probabilistic shortcut allows to reduce the complexities
  respectively to $O(n^{3+\varepsilon} \log (n))$ and
  $O(n^{4+\varepsilon})$.  Compared to the previous known attack due
  to Faure and Minder, our attack is efficient on codes from curves of
  arbitrary genus.  Furthermore, we investigate how far these methods
  apply to subcodes of AG codes.
\end{abstract}

\section*{Introduction}
\label{Introduction}
Most of the commonly used public key cryptosystems
% The security of nearly all public-key cryptosystems (PKC) - the ones that do not require an initial exchange of secrets -
are based only on two problems: the hardness of factoring or the
presumed intractability of the discrete logarithm problem.  However,
nothing ensures that the intractability of these problems will remain
true for the foreseeable future. In particular, we should notice that
all these classical number theory problems would be broken through
P.~Shor's quantum factorization algorithm~\cite{Shor:1994} in the case
a quantum computer would come to exist.  Thus, the cryptographic
community should look for alternative cryptosystems namely
\emph{Post-quantum cryptography}. Code-based Cryptography, together
with lattice-based cryptography, multivariate cryptography and
hash-based cryptography are the principal available techniques for
Post-quantum cryptography (for instance see \cite{bernstein:2009b}).

In the late seventies, McEliece \cite{mceliece:1978} introduced the first code based public-key cryptosystem
whose security reposes on the hardness of decoding a random linear code.
% or equivalently, the problem of finding minimum-weight codewords in a large linear code
Compared to public-key schemes based on integer factorization (like RSA)
or discrete logarithm, McEliece not only is resistant, so far, to attacks
by quantum computers, but also presents faster encryption and decryption
schemes. However, 
due to the large size of the keys required to have a good security level, it is rarely used in practice.
Nevertheless, note that recent proposals based on quasi--cyclic MDPC codes
\cite{misoczki:2012} allow compact keys of around 10000 bits for 128
bits of security, which makes such proposal competitive with RSA.

The original proposal \cite{mceliece:1978}, which remains unbroken, was based on binary Goppa codes. Later, several alternatives families with a higher correction capacity were proposed in order to reduce the key size. For instance, Generalized Reed-Solomon codes \cite{niederreiter:1986}, subcodes of them \cite{berger:2005} and Binary Reed-Muller codes \cite{sidelnikov:1994}, (note that this list is not exhaustive). All of these schemes are subject to polynomial or sub-exponential time attacks \cite{minder:2007, sidelnikov:1992, wieschebrink:2010}.

Another attempt, suggested by Janwa and Moreno \cite{janwa:1996} was to use algebraic geometry (AG) codes, their subfield subcodes or concatenated
AG codes. Take notice that:
\begin{itemize}
\setlength{\itemsep}{-.2mm}
\item The case of codes on curves of genus $0$ was broken by
  Sidelnikov and Shestakov \cite{sidelnikov:1992}.  For curves of
  genus $1$ and $2$ it was broken by Faure and Minder
  \cite{faure:2008}, but this attack has several drawbacks which makes
  it impossible to generalize to higher genera. For instance, the
  curve is required to be hyperelliptic, which is non generic for $g>
  2$ and the attack involves the computation of minimum-weight
  codewords whose cost is exponential in the genus of the curve.

\item In \cite{sendrier:1994}, Sendrier pointed out the inherent weakness
of concatenated codes for public key cryptography. Thus any proposal
using concatenation should be avoided.

\item In \cite{marquez:2013b}, the authors proved that the structure
  of the curve can be recovered from the only knowledge of a generator
  matrix of the code. Unfortunately, the efficient construction of a
  decoding algorithm from the obtained code's representation is still
  lacking. Thus, this result does not lead to an efficient attack.
\end{itemize}

In this article, we use another approach to attack the McEliece scheme
based on AG codes. Our attack is inspired by the attacks developed in
\cite{CGGOT12, COT14, COTG15, COT17} called {\em filtration attacks}. Such
attack uses the fact that the computation of some Schur products
  permits to distinguish AG codes from random ones. Moreover, this
  distinguisher can be used to compute an interesting filtration of
  the code used as a public key.  Compared to the previous
  filtration attacks which allowed to recover completely the structure
  of the public key, the present attack is not actually a key recovery
  attack. In particular, we do not compute the structure of the curve
  and the divisors providing the public code and we show that such
  computations are not necessary. Indeed, it is possible to stay in the
  realm of $\mathbb F_q^n$ and its subspaces in order to compute all
  the necessary data to get an efficient decoding algorithm for the AG
  code used as a public key. More precisely, we show how to use
  filtration attack techniques in order to compute an {\em
    error correcting pair} (see \cite{pellikaan:1988,
    pellikaan:1996}) allowing to correct up to $\frac{d^*-1-g}{2}$
  errors, where $d^*$ denotes the designed distance.
  These techniques can be pushed forward in order to compute
  an {\em error correcting array} \cite{kirfel:1995} or equivalently a
  {\em well behaving sequence} \cite{geil:2013}
  allowing to correct up to $\frac{d^*-1}{2}$ errors.
  The cost of this reconstruction is in $O(n^4 \log (n))$ operations on
  the base field
  for the reconstruction of an error correcting pair and
  $O(n^5)$ for the reconstruction of an error correcting array.
  A probabilistic shortcut permits one to reduce these complexities to
  $O(n^{3+\varepsilon})$ and $O(n^{4+\varepsilon})$ respectively.
  Finally, it is worth noting that many computations done in this filtration
  attack are very similar to those presented in a very different context by
  Khuri--Makdisi in \cite{khuri2001} in order to perform effective
  computations on Jacobians of curves.

  { The attack presented in this article is proved to be
    efficient for almost any algebraic geometry code. It should be
    emphasized that for some codes on curves of large genus $g$ and
    whose length satisfies $2g < n < 6g$, the proofs of efficiency of
    the attack do not hold.  Let us emphasize that this does not mean
    that the attack will fail for such codes but only that we have no
    mathematical proof for the efficiency of such cryptanalysis
    methods.  On the other hand, we show that there are several ways
    to extend the attack and we doubt that it would be possible to
    provide a large family of codes for which our attack would be
    guaranteed to fail.}

\paragraph{Outline of the article.}
% The reader is expected to be familiar with algebraic curves and AG codes
% which is for instance presented in \cite{TVN, stichtenoth:2009}.
Section~\ref{Section1} lists notation used in
the article and introduces some necessary material for the attack.
Next, we deal with two operations on codes: % and its
% applications concerned with cryptanalysis of public key cryptosystems:
the \emph{Schur product} introduced in Section~\ref{Section2} and a
new operation which we have called the \emph{{$s$}-closure}, defined in
Section~\ref{Section3}. The first one will be essential to attack
the McEliece scheme based on AG
codes while the second one will be crucial to attack
McEliece scheme based on subcodes of AG
codes.
Section~\ref{Section4} is devoted to the notion of
\emph{error-correcting pairs} (ECP) and \emph{error-correcting arrays}
(ECA) which provide efficient decoding algorithms for AG codes.
Section~\ref{Section5}
provides a detailed exposition of all the results needed for our
attack. In particular we show how to compute
an error correcting pair or an error correcting array
of a given code only by computing Schur products and performing Gaussian
elimination.
% Take notice
% that an ECA permits to correct more errors than an ECP.
% Therefore, the choice of
% computing one or the other option depends on the number of errors $t$
% that the cryptosystem admits, which is part of the public key.
Finally, Section~\ref{Section6} indicates how all the previous techniques may
be used to create an attack of the McEliece scheme based on AG codes
and Section~\ref{Section7} deals with the case of subcodes of AG
codes.

\paragraph{Note.} A part of the material of this article was presented
at the conferences ISIT\footnote{IEEE International Symposium
on Information Theory} 2014 and ICMCTA\footnote{International
Castle Meeting on Coding Theory and their Applications} 2014 and published in
 \cite{couvreur:2014a,couvreur:2014b}.
The present article provides a long version including detailed proofs which
were absent in the proceedings due to space reasons. It also includes
new results since the proceedings articles only considered the reconstruction
of error correcting pairs while we discuss here the reconstruction of error
correcting arrays.

%%% Local Variables:
%%% mode: latex
%%% TeX-master: "Long-Version"
%%% End:

\section{Prerequisites on curves and algebraic geometry codes}
\label{Section1}
%\label{Section1}
%\subsection{Curves and AG codes}

This section contains a brief summary of algebraic curves and
algebraic geometry (AG) codes to set up notation and
terminology. For a fuller treatment we refer the reader to
\cite{stichtenoth:2009, TVN}.

\subsection{Curves}
Let $\X$ denote a smooth projective geometrically connected curve over a finite field $\F_q$ and let $g$ denote the genus of $\X$.
The function field of the curve $\X$ with field of constants $\mathbb F_q$
is denoted by $\F_q(\X)$ and its elements are called \emph{rational functions}. Given a place $P$ of $\F_q(\X)$,
its degree is denoted as $\deg P$ and the \emph{valuation} at $P$
of any $f\in \F_q(\X)^{\times}$, is denoted by $\mathrm v_P(f)$.
%measures the order of zero or pole of $f$ at $P$.  Therefore, if $\mathrm v_P(f) = m>0$, then $P$ is a \emph{zero} of $f$ of multiplicity (or order) $m$ and if $\mathrm v_P(f) = m< 0$, then $P$ is a \emph{pole} of $f$ of order $-m$.
We use the convention $\mathrm v_p(0) = \infty$.
% In terms of the curve, a  place is a collection of points of $\X (\bar{\F}_q)$ that is an orbit under the action of the Frobenius map on the curve.
% Then, $\deg(P)$, the degree of the place is the number of points of
%  $\X (\bar{\F}_q)$ in this orbit.

A \emph{divisor} $E$ on $\X$ is a formal sum 
of places $E = \sum_{P\in \X} n_P P$ with $n_P \in \mathbb Z$.
The \emph{degree} of $E$ is the integer $\deg(E) \eqdef 
\sum_{P\in \X} n_P\deg(P)$ and the {\em support} of $E$, is the set of places with $n_P \neq 0$.
%\todo{I removed notation supp which is never used after.}
If all coefficients $n_P$ are nonnegative, $E$ is an \emph{effective} divisor,
denoted by $E\geq 0$. Next, we denote by $E \geq F$ when $E-F \geq 0$.
This defines a partial order on the group of divisors.

Let $f\in \F_q(\X)\setminus \{0\}$, the divisor of $f$ is denoted by $(f)$.
% , as
% $$(f) = \sum_{P \hbox{ zero of } f} \mathrm v_P(f) P + \sum_{P \hbox{ pole of } f }\mathrm v_P(f)P.$$
% Therefore, we think of $(f)$ as \emph{``the zeros of $f$ minus the poles of $f$''}.
Given a divisor $E$ on $\mathcal X$, the corresponding \emph{Riemann Roch space} is denoted by $L(E)$ and is defined as % the space of rational functions with zeroes and poles prescribed by $E$ as
follows
$$
L(E) = \left\{ f\in \F_q(\X) \mid f = 0 \hbox{ or } (f) + E \geq 0\right\}.
$$

%\todo[inline]{Section E-gap from \S 2.4 to here, since this is purely AG and now the definition of $\hat{r}(i,j)$ comes after the one of $r(i,j)$}
\subsection{Weierstrass gaps}
\label{Weierstrass-gaps}
Gaps and non gaps are fundamental notions in this article. For this reason
we remind several very usual properties of these objects. 
See \cite{kirfel:1995} for further details.

\begin{definition}
\label{GAP}
Let $E$ be a divisor on $\X$ and $P$ be a rational point.
A positive integer $i$ is called an $E$-{\em gap} at $P$ if $L(E+iP) = L(E+(i-1)P)$. Otherwise, $i$ is an $E$ {\em non-gap} at $P$.
In case $E=0$, we just say \emph{gap} and \emph{non-gap}, respectively.
\end{definition}

% \begin{remark}
% \label{Remark::Gap}
%In the following we provide some basic properties of the set of gaps.
The following statement lists many elementary properties of gap
and non-gap sets.
\begin{proposition}\label{prop::Gap}
\begin{enumerate}[$(1)$]
\setlength{\itemsep}{-.2mm}
\item \label{Gap:1} % By Theorem \ref{Theorem::1} 
  If $i\geq -\deg(E) + 2g$ then $i$ is an $E$ non-gap at $P$.
\item \label{Gap:2} The $E$-gaps at $P$ lie in the interval
$[-\deg(E), -\deg(E) + 2g-1]$ and the number of gaps is exactly equal to $g$.
% The first statement is a reformulation of (\ref{Gap:1}). For the second one, take notice the fact that the dimension of $L(E+iP)$ increases with at most one if we increase $i$ by one, i.e. $$\dim (L(E + (i-1)P)) \leq \dim(L(E+iP)) \leq \dim (L(E+(i-1)P))+1.$$
% Moreover, $\dim(L(E+iP)) = 0$, if $i=-\deg(E)$, and $\dim(L(E+iP)) =g$, if $i=-\deg(E) + 2g-1$.
\item \label{Gap:3} If $\alpha$ is an F non-gap at $P$ and $\beta$ is an $E$ non-gap at $P$. Then, $\alpha + \beta$ is an $(F+E)$ non-gap at $P$.
In particular if $\alpha$ is a non-gap at $P$ and $\beta$ is an $E$ non-gap at $P$. Then, $\alpha+\beta$ is an $E$ non-gap at $P$.
\item \label{Gap:4}
Let  $(\alpha_i)_{i\in \mathbb N}$ be the non-gap sequence at $P$ and $(\beta_j)_{j\in \mathbb N}$ be the $E$ non-gap sequence at $P$.
Then,
$$
j = \dim L(E+\beta_j P) = \dim L(E+(\beta_{j-1})P) + 1.
$$
Thus, $j-1 \leq \deg(E)+\beta_j \leq  j +g-1$. Moreover,
$\deg(E)+\beta_j = j +g-1$ if $\deg(E) + \beta_j > 2g-1$, i.e. $j> g$.
Similarly $i-1 \leq \alpha_i \leq  i +g-1$. Thus, $\alpha_i = i +g-1$ if $i> g$.

% \item \label{Gap:5} For all positive integers $i,j$ there is a unique $r = r(i,j)$ such that $\alpha_i + \beta_j = \beta_r$. % \\
% % Since $\alpha_i + \beta_j$ is an $E$ non-gap at $P$ by Remark \ref{prop::Gap}-(\ref{Gap:3}).

% \item \label{Gap:5a} The function $r(i,j)$ is strictly increasing in
% both arguments, since the sequences $(\alpha_i)_{i\in \mathbb N}$
% and $(\beta_j)_{j\in \mathbb N}$  are strictly increasing.
\end{enumerate}
\end{proposition}

\subsection{Algebraic geometry codes}
\label{sec:AG_codes}
{We assume that the reader is aware of basic notions of
  coding theory and refer to~\cite{MS86} for further details.
  Below we remind some basic
  notions on algebraic geometry codes and refer the reader to
  \cite{TVN, stichtenoth:2009}.}

Given an $n$--tuple $\P = (P_1, \ldots, P_n)$
of pairwise distinct $\F_q$--rational points of $\X$, we denote by $\DP$
the divisor $\DP \eqdef P_1+\cdots +P_n$. Let $E$ be a divisor of $\X$ with disjoint support from $P$, then the evaluation map
$$
\ev_{\P} : 
\map{L(E)}{\F_q^n}{f}{(f(P_1), \ldots, f(P_n))}
$$
is well--defined.
% by $\ev_{\P}(f) = \left( f(P_1), \ldots, f(P_n)\right)$.

\begin{definition}
Let $\P=(P_1, \ldots, P_n)$ be an $n$-tuple of mutually distinct $\F_q$-rational points of the curve $\X$ and let $E$ be a divisor of $\X$ with disjoint support from $\DP$. Then, the \emph{algebraic geometry} (AG) code $\C{X}{P}{E}$ of length $n$ over $\F_q$ is the image of $L(E)$ under the evaluation map $\ev_{\P}$, that is
$$
 \C{X}{P}{E} = \left\{ \ev_{\P}(f) \mid f\in L(E)\right\}.
$$
\end{definition}

From now on, the dimension of a linear code $C$ will be denoted by $k(C)$ and its minimum distance by $d(C)$. Let $\X$, $\P$ and $E$ be respectively a smooth projective
geometrically connected curve over $\F_q$, an $n$--tuple of $\F_q$--rational
points of $\X$ and a divisor on $\X$. 
Let us remind some well--known statements.

\begin{theorem}[{\!\!\cite[Theorem 2.2.2]{stichtenoth:2009}}]
\label{Theorem::1}
If $\deg(E) < n$, then
\begin{align*}
  k\left(\C{X}{P}{E}\right) &\geq \deg(E) + 1 -g \\
  d\left(\C{X}{P}{E} \right) &\geq n-\deg(E).
\end{align*}
% $$\begin{array}{ccc}
% k\left(\C{X}{P}{E}\right)\geq \deg(E) + 1 -g & \hbox{ and }&
% d\left(\C{X}{P}{E} \right) \geq n-\deg(E).
% \end{array}$$
Moreover, if $n>\deg(E) > 2g-2$, then $k(\C{X}{P}{E})=\deg(E) - g +1$.
\end{theorem}

%The dual of an AG code is an AG code and we have the following result.

\begin{theorem} [{\!\!\cite[Proposition 2.2.10]{stichtenoth:2009}}]
\label{Theorem::2}
There exists a differential form $\omega$ with a simple pole and residue
$1$ at $P_j$ for all $j \in \{1, \ldots, n\}$.
Let $K$ be the divisor of $\omega$, then
$$\C{X}{P}{E}^{\perp} = \C{X}{P}{E^{\perp}},$$
where $E^{\perp} \eqdef  \DP-E+K$ and $\deg(E^{\perp}) = n-\deg(E)+2g-2$.
\end{theorem}

%\todo{paragraph deleted}

% \begin{corollary}
% \label{Corollary:Dual}
% If $\deg(E)>2g-2$, then
% \begin{align*}
%   k\left(\C{X}{P}{E}^{\perp}\right) & \geq  n-\deg(E)-1+g\\
%   d\left(\C{X}{P}{E}^{\perp}\right) & \geq \deg(E)-2g+2.
% \end{align*}
% % $$\begin{array}{ccc}
% % k\left(\C{X}{P}{E}^{\perp}\right) \geq  n-\deg(E)-1+g & \hbox{ and \ \ }
% % d\left(\C{X}{P}{E}^{\perp}\right) \geq \deg(E)-2g+2.
% % \end{array}$$
% Moreover, if $n>\deg(E)>2g-2$, then $k\left(\C{X}{P}{E}^{\perp}\right)
% =n-\deg(E)-1+g$.
% \end{corollary}
% \todo[inline]{I removed the corollary on the parameters of dual codes. It was cited only once and the citation was avoidable}

\section{Schur product of codes}
\label{Section2}
The notion of Schur product of codes was first introduced in coding theory
for decoding \cite{pellikaan:1988}. Next,
this apparently trivial operation turned out to have many other applications
such as cryptanalysis, multiparty computation, secret sharing
 or construction of lattices.
Many of these applications are summarized in \cite[\S 4]{Hugues}.
 
\begin{definition}
The Schur product is the component wise product on $\F_q^n$:
given two elements $\ab$ and $\bb$ in $\Fq^n$,
$$
\ab * \bb \eqdef (a_1b_1, \ldots, a_nb_n).
$$
% Let $\ab \in \Fq^n$, we set $\ab^0 = (1,\ldots , 1)$
% and by induction we define $\ab^{j+1} = \ab * \ab^j$ for any
% positive integer $j$. If all entries of $\ab$ are nonzero,
% we define $\ab^{-1} = (a_1^{-1}, \ldots , a_n^{-1})$ and thus, $\ab^{-j} = (a_1^{-j},
% \ldots, a_n^{-j})$ for any positive integer $j$.
% \todo{I remove notation $\ab^j$ which is (I think) never used in what follows}
For two codes $A, B \subseteq \Fq^n$, their Schur product is the code $A*B$
defined as
$$A*B \eqdef  \Span_{\Fq} \left\{ \ab * \bb  \mid \ab \in A \hbox{ and } \bb \in
B\right\}.$$
For $B=A$, then $A*A$ is denoted as $A^{(2)}$ and, we define $A^{(t)}$
by induction for any positive integer $t$.
%Furthermore, we denote by $A \perp B$ if
%$\ab \cdot \bb = 0$ for all $\ab \in A$ and $\bb \in B$.  
\end{definition}

% The Schur product has applications in many different contexts such as algebraic decoding,
% cryptanalysis, multiparty computation, secret sharing, oblivious transfer or construction of lattices.
% Many of these applications are summarized in \cite[\S 4]{Hugues} and
% this paper gives further details on the applications to decoding and to cryptanalysis.

\subsection{Relation with the canonical inner product}
Take notice that the Schur product
should not be confused with the \emph{standard inner product} which
is defined as
$$
\langle \ab, \bb \rangle \eqdef  \sum_{i=1}^n a_ib_i.
$$
However, the two notions are related by the following elementary
adjunction principle:
%\begin{lemma}
 $$ \forall \ab, \bb, \cb \in \F_q^n,\quad
 \langle \ab * \bb , \cb \rangle = \langle \ab , \bb * \cb\rangle
 = \sum_{i=1}^n a_ib_ic_i.$$
%\end{lemma}
An interesting consequence of this relation between the Schur product
and the canonical inner product is the following statement which is very
useful in what follows.

\begin{lemma}
\label{Lemma::9}
Let $A$ and $B$ be two codes in $\F_q^n$. Then
$$\left\{ \mathbf z\in \F_q^n \mid \mathbf z * A \subseteq B\right\} = \left(
A * B^{\bot} \right)^{\bot}.$$
\end{lemma}

\begin{proof}
It is easily seen that
\begin{eqnarray*}
\mathbf z * A \subseteq B & \Leftrightarrow &
\langle \mathbf z * \mathbf a, \mathbf b\rangle = 0
\hbox{ , } \forall \mathbf a \in A \hbox{ and } \forall \mathbf b \in B^{\bot}\\
 & \Leftrightarrow & \langle \mathbf z,  \mathbf a * \mathbf b \rangle = 0
\hbox{ , } \forall \mathbf a \in A \hbox{ and } \forall \mathbf b \in B^{\bot} \Leftrightarrow  \mathbf z \in \left(A*B^{\bot}\right)^{\bot}.
\end{eqnarray*}
\end{proof}

\subsection{Schur product of algebraic geometry codes}

An interesting aspect of the Schur product is that the evaluation map
$\ev_{\mathcal{P}}$ introduced in Section~\ref{sec:AG_codes} arises
from a morphism of algebras.
More precisely, let $\mathcal O_{\mathcal{P}}$ be the subring of $\F_q(\X)$
of functions regular at $P_1, \ldots, P_n$ then, the map
$\ev_{\mathcal{P}} : \mathcal{O}_{\mathcal{P}} \rightarrow \F_q^n$ is a morphism
of algebras since for all $f, g \in \mathcal{O}_{\mathcal{P}}$,
we have $\ev (fg) = \ev (f) * \ev (g)$.
Therefore, many arithmetic properties of the function field $\F_q (\X)$
can be understood in terms of AG codes thanks to the Schur product.

To understand the behaviour of Schur products of AG codes, we need to
analyze a similar operation in terms of the function field.  That is,
we need to understand the behaviour of spaces
defined by products of elements of two finite dimensional subspaces of
a given function field.  Let us first introduce another notation: let
$\mathbb A$ be a commutative unitary algebra over a field
$\mathbb{K}$.  Given two subspaces $V, W$ of $\mathbb{A}$, in the same
way as for the Schur product, the product of $V$ and $W$ is defined
as:
 $$
 V \cdot W \eqdef  \Span_{\mathbb{K}} \left\{ vw ~|~ v\in V,\ w\in W\right\}.
 $$
We define inductively $L^{(1)}=L$ and $L^{(t+1)} = L^{(t)}\cdot L$.

\begin{theorem}[{\!\!\cite[Theorem 6]{mumford:1970}}]
\label{Theorem::6}
Let $E, F$ be two divisors on the curve $\X$ such that $\deg(E) \geq 2g+1$ and $\deg(F) \geq 2g$ and let $t$ be a positive integer. Then,
\begin{enumerate}[$(1)$]
\setlength{\itemsep}{-.2mm}
 \item\label{thm6-(1)} $L(E) \cdot L(F) = L(E+F)$;
 \item \label{thm6-(2)} $L(E)^{(t)} = L(tE)$.
\end{enumerate}
\end{theorem}

Since the evaluation map is a morphism of algebras we deduce directly
from Theorem~\ref{Theorem::6} the following statement.

\begin{corollary}
\label{Corollary::7}
Let $E, F$ be two divisors on the curve $\X$ both with disjoint support
with $\P$ and such that $\deg(E) \geq 2g+1$ and $\deg(F)\geq 2g$
and let $t$ be a positive integer. Then,
\begin{enumerate}[$(1)$]
\setlength{\itemsep}{-.2mm}
 \item\label{cor7-(1)}  $\C{X}{P}{E} * \C{X}{P}{F} = \C{X}{P}{E+F}$;
\item \label{cor7-(2)}  $\C{X}{P}{E}^{(t)} = \C{X}{P}{tE}$.
\end{enumerate}
\end{corollary}

% \begin{proof}
% The evaluation map is a morphism of algebras,
% since it is linear and $\ev_{\P}(f.g) = \ev_{\P}(f) * \ev_{\P}(g)$.
% So the proposition is a direct consequence of Theorem \ref{Theorem::6}.
% \end{proof}

We conclude this subsection with the following statement
which is crucial in the sequel.

\begin{proposition}
  \label{Proposition::8-9}
  Let $E, F$ be two divisors on the curve $\mathcal X$ both with
  disjoint supports with $\P$ and such that $\deg(F) \geq 2g$ and $\deg(E)
  \leq n-3$. Then,
\begin{align*}
 \C{X}{P}{E-F} & = \left( \C{X}{P}{F} *\C{X}{P}{E}^{\bot} \right)^{\bot} \\
 & =   \left\{ \mathbf z \in \F_q^n \mid \mathbf z * \C{X}{P}{F} \subseteq
\C{X}{P}{E} \right\}.
\end{align*}
% the following spaces are equal:
% \begin{enumerate}[(i)]
%   \item $\left( \C{X}{P}{F} *\C{X}{P}{E}^{\bot} \right)^{\bot} $;
%   \item $\left\{ \mathbf z \in \F_q^n \mid \mathbf z * \C{X}{P}{E} \subseteq
% \C{X}{P}{F} \right\}$;
%   \item $\C{X}{P}{E-F}$.
% \end{enumerate}
\end{proposition}
%\todo{I merged two statements.}

\begin{proof}
  Let $A = \C{X}{P}{F}$ and $B = \C{X}{P}{E}$.
Theorem \ref{Theorem::2} shows that
$B^{\bot} = \C{X}{P}{E^{\bot}}$ %for some $E^{\bot}$
with $E^{\bot} = K- E + \DP$ for some canonical divisor $K$.
Next, since $\deg(E) \leq n-3$, we have
$\deg (E^{\bot})  \geq 2g +1$.
Now,
$$A * B^{\bot} = \C{X}{P}{\DP + K - E + F} = \C{X}{P}{E-F}^{\bot},$$ 
the first equality being a consequence of Corollary~\ref{Corollary::7} and the last one due to Theorem \ref{Theorem::2}.
The second equality of the statement follows from Lemma \ref{Lemma::9}.
\end{proof}

% \begin{proposition}
% \label{Proposition::8}
% Let $E, F$ be two divisors on the curve $\mathcal X$ with disjoint support with $\DP$ such that $\deg(F) \geq 2g$ and $\deg(E) \leq n-3$. Then,
% $$\left( \C{X}{P}{F} *\C{X}{P}{E}^{\bot} \right)^{\bot} = \C{X}{P}{E-F}$$
% \end{proposition}

% \begin{proof}
% Let $A = \C{X}{P}{F}$ and $B = \C{X}{P}{E}$. Since $\deg(E) \leq n-3$,
% Theorem \ref{Theorem::2} shows that
% $$\begin{array}{ccc}
% B^{\bot} = \C{X}{P}{E^{\bot}} & \hbox{ with } &
%  \deg(E^{\bot}) = n-\deg(E) + 2g - 2 \geq 2g +1
% \end{array}$$
% where $E^{\bot} = \DP + K - E$ for some canonical divisor $K$ on $\X$.

% Now,
% $A * B^{\bot} = \C{X}{P}{\DP + K - E + F} = \C{X}{P}{E-F}^{\bot} $, the first equality being a consequence of Corollary~\ref{Corollary::7} and the last one due to Theorem \ref{Theorem::2}.
% \end{proof}

% \begin{proposition}
% \label{Proposition::10}
% Let $E, F$ be two divisors on the curve $\X$ with disjoint support with $\DP$ such that $\deg(F) \geq 2g$ and $\deg(E) \leq n-3$. Then,
% $$\left\{ \mathbf z \in \F_q^n \mid \mathbf z * \C{X}{P}{E} \subseteq
% \C{X}{P}{F} \right\} = \C{X}{P}{E-F} $$
% \end{proposition}

% \begin{proof}
% The statement follows from Lemma \ref{Lemma::9} and Proposition \ref{Proposition::8}.
% \end{proof}

\subsection{Distinguisher and cryptanalysis}

Another and more recent application of the Schur product concerns
cryptanalysis of code-based public key cryptosystems. In this context,
the Schur product is a very powerful operation which can help to
distinguish some algebraic codes such as AG codes from random ones.
The point is that evaluation codes do not behave like random codes
with respect to the Schur product: the square of an AG code is 
small compared to that of a random code of the same dimension. Thanks
to this observation, Wieschebrink \cite{wieschebrink:2010} gave an
efficient attack of Berger Loidreau's proposal \cite{berger:2005}
based on subcodes of GRS codes.

Recent attacks consist in using this argument and take
advantage of this distinguisher in order to
compute a filtration of the public code by a family of very particular
subcodes.  This filtration method yields an alternative attack on GRS
codes \cite{CGGOT12}. Next it lead to a key recovery attack on wild
Goppa codes over quadratic extensions in \cite{COT14, COT17}.  Finally in the
case of AG codes, this approach leads to an attack as we will see in
Section~\ref{Section6}. This attack consists in the computation of an
error-correcting pair (ECP) or an error-correcting array (ECA) for the
public code without retrieving the structure of the curve, the points
and the divisor.

\section{The {$s$}-closure operation}\label{sect:t-closure}
\label{Section3}
In this section we introduce a new operation which we call the
\emph{{ $s$-}closure}. This operation will be crucial in Section~\ref{Section4} to attack a McEliece scheme based on subcodes of AG
codes.
Roughly speaking, given a random subcode $C$ of an AG code
$\C{X}{P}{E}$, if the codimension of $C$ in
$\C{X}{P}{E}$ is {\em small enough}, then, with a high probability,
the $2$-closure of $C$ provides the code $\C{X}{P}{E}$.

% Roughly speaking this operation allows us to recover the AG code
% from which the public code is a subcode. Indeed, suppose that our
% public key is a non structured generator matrix $G$ of a subcode
% $\mathcal C$ of $\C{X}{P}{E}^{\perp}$ of dimension $l$, together
% with the error correcting capacity $t$. We will show that the
% $2$-closure of $\mathcal C$ provides the code $\C{X}{P}{E}$. In
% other words, if we have an attack that recovers a $t$-decoding
% algorithm for $\C{X}{P}{E}$ , then this attack will also work for
% the subcode $\mathcal C$.

\begin{definition}[{ $s$--}closure]
  Let $\mathbb{A}$ be a commutative unitary algebra over a field
  $\mathbb{K}$ and suppose that we have a subspace $L$ of $\mathbb A$.
  { Let $s \geq 2$ be an integer,} the {
    $s$--}{\em closure} of $L$ is defined by
   $$
   \overline{L}^{s} = \left\{ f \in \mathbb A \mid f \cdot
     L^{({s}-1)} \subseteq L^{({s})}\right\}.
   $$
   The space $L$ is called { $s$--}{\em closed} if
   $\overline{L}^{s} = L$.
\end{definition}

\begin{remark}
  Special cases are discussed in the sequel where $\mathbb{K} = \Fq$
  and either $\mathbb A$ is the field of rational functions on the
  curve $\X$ and $L$ is a subspace of a Riemann-Roch space $L(E)$ for
  a divisor $E$ on the curve,
  or $\mathbb A = \Fq^n$ and $L$ is a subspace of $\Fq^n$.
\end{remark}

\subsection{General properties}
We list below some properties of the { $s$-}closure operation.
\begin{proposition}
  \label{Proposition::11}
  Let $L,~M$ and $N$ be subspaces of $\mathbb A$ { and $s \geq 2$
  be an integer}, then:
  \begin{enumerate}[$(1)$]
    \setlength\itemsep{-.2mm}
  \item \label{it:1} $\overline{L}^{s}$ is a vector space
    over $\mathbb{K}$.
    \item \label{it:2} We have the following increasing sequence: $L
      \subseteq \overline{L}^2 \subseteq \cdots \subseteq
      \overline{L}^{s}\subseteq \overline{L}^{{s}+1}$.
      % \item \label{it:3}Let $f\in \mathbb A$. Then $ (f\cdot L)^{(t)} = f^t
      %   \cdot L^{(t)}$.
    \item \label{it:4} Let $f\in \mathbb A$. Then
      $ f\cdot \overline{L}^{s} \subseteq \overline{(f\cdot
        L)}^{s}$.
      Equality holds if $f$ is invertible in $\mathbb{A}$.
    \item \label{it:5} Let $\phi : \mathbb{A} \rightarrow \mathbb{K}$
      be a linear form and consider the non degenerate symmetric
      bilinear form over $\mathbb{A}$ defined as $\varphi(a,b) \eqdef
      \phi(ab)$. Then
      $$
      \overline{L}^{s} = {\left( L^{({s}-1)}
          \cdot {L^{({s})}}^{\bot_{\varphi}}
        \right)}^{\bot_{\varphi}}.
      $$
      This holds in particular when $\mathbb{A} = \Fq^n$ and $\varphi$
      is the standard inner product.
 \end{enumerate}
\end{proposition}

\begin{proof}
% \begin{enumerate}[$(1)$]
%\setlength\itemsep{0em}
Statement (\ref{it:1}) follows from the $\mathbb{K}$-bilinearity of the product.
To prove  (\ref{it:2}), let $f \in \overline{L}^{s} $. Thus,
$$
f \cdot L^{({s}-1)} \subseteq L^{({s})} \ \ \Longrightarrow \ \
f \cdot L^{({s})} \subseteq L^{({s}+1)}.
$$
Therefore, $f \in \overline{L}^{{s}+1} $.
% The space $(f \cdot L)^{(t)}$ is generated by elements $g$ of the form
% $\prod_{i=1}^t ff_i$ with $f_i \in L$ for all $i$. Now $\prod_{i=1}^t
% ff_i= f^t\prod_{i=1}^tf_i$ and $\prod_{i=1}^tf_i\in L^{(t)}$.  So $g
% \in f^t \cdot L^{(t)}$. Hence $ (f\cdot L)^{(t)} = f^t \cdot L^{(t)}$
% by linearity
To prove (\ref{it:4}), let $g \in f\cdot \overline{L}^{s} $. Then,
$g = fa$ for some $a \in  \overline{L}^{s}$ and
$$
g\cdot {(f\cdot L)}^{({s}-1)} = f^{s} \cdot a
\cdot L^{({s}-1)} \subseteq f^{s} \cdot
L^{({s})} = (f\cdot L)^{({s})}
$$
% $f^{-1}g \in \overline{L}^t $.  Thus, $f^{-1}g \cdot L^{(t-1)}
% \subseteq L^{(t)}$, or equivalently, $f^t\cdot f^{-1}g \cdot L^{(t-1)}
% \subseteq f^t \cdot L^{(t)}$.  So $g \cdot (f\cdot L)^{(t-1)}
% \subseteq (f \cdot L)^{(t)}$ by (\ref{it:3}). 
Therefore $g \in \overline{f\cdot L}^{s}$.  If moreover $f$
is invertible, then one can reverse the proof and the equality holds.
Finally the proof of (\ref{it:5}) is in the very same spirit as that
of Lemma~\ref{Lemma::9} using the adjunction formula
$\varphi (fg,h)= \phi(fgh)= \varphi (f,gh)$, which holds for all
$f,g,h \in \mathbb{A}$.
% To prove (\ref{it:5}), notice that .
% Hence the following statements are equivalent:
% \begin{itemize}
% \item[-] $f \in \overline{L}^t $
% \item[-] $fg \in L^{(t)} $ for all $g \in L^{(t-1)} $
% \item[-] $\varphi (fg,h)=0$ for all $g \in L^{(t-1)} $ and for all $h \in (L^{(t)})^{\bot_{\varphi}} $
% \item[-] $\varphi (f,gh)=0$ for all $g \in L^{(t-1)} $ and for all $h \in (L^{(t)})^{\bot_{\varphi}} $ (by linearity)
% \item[-] $\varphi (f,u)=0$ for all $u \in L^{(t-1)} \cdot (L^{(t)})^{\bot_{\varphi}} $
% \item[-] $f \in \left( L^{(t-1)} \cdot (L^{(t)})^{\bot_{\varphi}} \right)^{\bot_{\varphi}}$
% \end{itemize}
%\end{enumerate}
%\todo{I shortened the proof.}
\end{proof}

Notice that one can have $L\subseteq M$ while $\overline{L}^{s} \not
\subseteq \overline{M}^{s}$
as illustrated by the following example.

   \begin{example}
     \label{example::non_inclusion}
     Let $\mathbb A = \mathbb{K} [x]$ and
     $
     L \eqdef \Span \{1, x + x^2,  x^3 , x^4\}
     $
     and
     $
     M \eqdef L \oplus \Span \{x^{9}\}.
     $
     A computation gives
     $$
     L^{(2)} = \mathbb K [x]_{\leq 8} \qquad {\rm and} \qquad
     M^{(2)} = \mathbb K [x]_{\leq 8} \oplus x^{9} \cdot L \oplus \Span \{x^{18}\},
     $$
     {where $\mathbb K [x]_{\leq 8}$ denotes the finite
     dimensional subspace of $\mathbb K [x]$ of polynomials of degree less
     than or equal to $8$.}
     Next, one proves easily that $x \in \overline L^2$ while $x
     \notin \overline M^2$ since $x\cdot x^9 = x^{10}\notin M^{(2)}$.
     Therefore, $\overline L^2 \not \subseteq \overline M^2$.
   \end{example}

   On the other hand, we have the following lemma.

 \begin{lemma}
 \label{Lemma::12}
 Let $\mathbb A$ be the field of rational functions on the curve $\X$,
 let $M=L(E)$ for a divisor $E$ on $\mathcal X$ {
 and $s \geq 2$ be an integer}.  If $L$ is a
 subspace of $M$, $L^{({ s})} = M^{({ s})}$ and
 $\overline{M}^{ s}=M$, then $\overline{L}^{ s} = M$.
 \end{lemma}

 \begin{proof}
 Assume that $L\subseteq L(E)$.
 Let $E_0=\sum a_iP_i$ be divisor satisfying $L\subseteq L(E_0) \subseteq L(E)$
 and minimal for this property.
 Then, for all $i$ there exists a function $g_i\in L$ such that $v_{P_i}(g_i)=-a_i$.
 Take $f \in \overline{L}^{ s}$. By definition,
 $$
 f\cdot g_i^{{ s}-1} \in f \cdot L^{({ s}-1)}
 \subseteq L^{({ s})} \subseteq L(E_0)^{({ s})}
 \subseteq L({ s}E_0).
 $$
 So $v_{P_i}(fg_i^{{ s}-1})=v_{P_i}(f)-({ s}-1)a_i\geq
 -{ s}a_i$.
 Or equivalently, $v_{P_i}(f) \geq -a_i$ for all $i$.  Therefore,
 $(f)\geq -E_0$, that is $f \in L(E_0)\subseteq L(E)=M$, and
 $\overline{L}^{ s} \subseteq M$ is proved.

 Conversely, let $f \in M$, we have $f \cdot L^{({ s}-1)} \subseteq f
 \cdot M^{({ s}-1)} \subseteq M^{({ s})} =L^{({ s})}$
 and hence $f \in \overline{L}^{ s}$.
 \end{proof}

\subsection{Closures of Riemann Roch spaces and AG codes}
We will show that certain Riemann-Roch spaces of AG codes are
{ $s$-}closed. For this sake we first need the following lemma.

\begin{lemma}
\label{Lemma::5}
Let $E, F$ be two divisors on the curve $\X$ of genus $g$ with $\deg(E) \geq 2g$. Then,
$$\begin{array}{ccc}
E\leq F & \hbox{ if and only if }& L(E) \subseteq L(F).
\end{array}$$
\end{lemma}

 \begin{proof}
% It is always the case that $L(E) \subseteq L(F)$ if $E \leq F$.
 The ``only if'' part is obvious.
 Conversely, let $L(E) \subseteq L(F)$ and suppose that $E\not\leq F$.
 If $E=\sum m_PP$ and $F=\sum n_PP$ with each $m_P, ~n_P \in \mathbb Z$, then, since $E \not\leq F$, there is a place $P_0$ such that $m_{P_0}>n_{P_0}$.
 As $\deg(E) \geq 2g$, by Riemann-Roch Theorem,
 $\dim (L(E)) = \deg(E)+1 -g >g$. We distinguish two different cases:
 \begin{itemize}
   \setlength{\itemsep}{-.2mm}
 \item If $\deg (E-P_0) >2g-2$, then $\dim (L(E-P_0))= \deg(E-P_0)+1 -g < \dim (L(E))$.
 \item If $\deg (E-P_0) \leq 2g-2$, then $\dim (L(E-P_0)) \leq \frac12 \deg(E-P_0)+1 $, by Clifford's Theorem. So $\dim(L(E-P_0))\leq g < \dim(L(E))$.
 \end{itemize}
 In both cases, $\dim(L(E-P_0)) < \dim (L(E))$.  Hence, there exists a
 rational function $f \in L(E)\setminus L(E-P_0)$.  That means that
 $v_{P_0}(f) = -m_{P_0}<-n_{P_0}$. So $f \not\in L(F)$, which
 contradicts our initial assumption.
 \end{proof}

% \begin{lemma}
%  \label{Lemma::13}
%  Let $E$ be a divisor on the curve $\X$ of genus $g$ with $\deg (E) \geq 2g+1$.
%  Let $i$ and $t$ be positive integers with $t> i$. Let $f$ be a nonzero rational function on $\X $. Then:
% $$\begin{array}{ccc}
% f\cdot L(E)^{(i)} \subseteq L(E)^{(t)} & \Longleftrightarrow & f\in L(E)^{(t-i)}
% \end{array}$$
%  \end{lemma}
%  \begin{proof}
% First notice that $f\cdot L(E) = L(E-(f))$.
%  Indeed,
%  \begin{eqnarray*}
%  h \in f\cdot L(E)
%  &\Leftrightarrow& hf^{-1} \in  L(E)
%  \Leftrightarrow (h)-(f) =(hf^{-1}) \geq -E\\
%  &\Leftrightarrow& (h)\geq -(E-(f))
%  \Leftrightarrow h \in L(E-(f))
%  \end{eqnarray*}

%  Now, let $f\in L(E)^{(t-i)}$, then $f\cdot L(E)^{(i)} \subseteq L(E)^{(t)}$ by definition of $L(E)^{(t)}$.\\
%  Conversely, suppose that $f\cdot L(E)^{(i)} \subseteq L(E)^{(t)}$. Then by Theorem \ref{Theorem::6}-(\ref{thm6-(2)}):
%  $$ L(iE-(f)) = f\cdot L(E)^{(i)} \subseteq  L(E)^{(t)} = L(tE).$$
%  Hence, $iE-(f)\leq tE$ by Lemma \ref{Lemma::5}, or equivalently, $(f) \geq -(t-i)E$.
%  Thus, we can conclude that $f\in L((t-i)E)=L(E)^{(t-i)}$ by Theorem \ref{Theorem::6}-(\ref{thm6-(2)}).
%  \end{proof}

 \begin{proposition}
 \label{Proposition::14}
 Let $E$  be a divisor on $\X$ with $\deg (E) \geq 2g+1$
 { and $s \geq 2$ be an integer}.
 Then, $\overline{L(E)}^{ s} =L(E)$.
 \end{proposition}
 \begin{proof}
 Inclusion $L(E) \subseteq  \overline{L(E)}^{ s}$ is obvious.
 Conversely, let $f \in \overline{L(E)}^{ s}$.
 By definition, %$f \in L(E)^{(t)}$ and
 $ f\cdot L(E)^{({ s}-1)} \subseteq  L(E)^{({ s})}$.
 From Theorem~\ref{Theorem::6}~(\ref{thm6-(2)}), this gives
 $f \cdot L(({ s}-1)E) \subseteq L({ s}E)$.
 A simple computation shows that $f \cdot L (({ s}-1)E) =
 L(({ s}-1)E - (f))$.
 Therefore, we have the inclusion 
 $L(({ s}-1)E - (f)) \subseteq L({ s}E)$ and,
 thanks to Lemma~\ref{Lemma::5}, we get
 $({ s}-1)E - (f) \leq { s}E$, which entails
 $(f) \geq -E$ and hence $f \in L(E)$.
 \end{proof}

In terms of AG codes, this leads to:

\begin{proposition}
\label{Proposition::15}
Let $E$ be a divisor on the curve $\X$ such that $2g+1 \leq \deg(E) \leq \frac{n-2}{t}$.
Let ${ s}$ be an integer with ${ s}\geq  2$.
Then,
$$\overline{\C{X}{P}{E}}^{ s} = \C{X}{P}{E}. $$
\end{proposition}

\begin{proof}
Proposition~\ref{Proposition::11}(\ref{it:5}) gives that
\begin{equation}\label{eq:closure}
\overline{\C{X}{P}{E}}^{ s} = \left(
\C{X}{P}{E}^{({ s}-1)} *
\left(\C{X}{P}{E}^{({ s})}\right)^{\perp}
\right)^{\perp}.
\end{equation}
Now, from Corollary~\ref{Corollary::7}(\ref{cor7-(2)}),
$\C{X}{P}{E}^{({ s})} = \C{X}{P}{{ s}E}$.
Moreover, by Theorem \ref{Theorem::2}, we have
$\C{X}{P}{{ s}E}^{\perp} = \C{X}{P}{({ s}E)^{\perp}}$
with $({ s}E)^{\perp} =
\DP - { s}E + K$ for some canonical divisor $K$ on $\mathcal{X}$. 
By assumption, $\deg(E) \leq \frac{n-2}{{ s}}$ and hence
$$\deg\left(({ s}E)^{\perp}\right) = n-\deg({ s}E)
+2g-2\geq 2g.$$
As $\deg(E) \geq 2g+1$, then, thanks to 
%\sout{Corollary~\ref{Corollary::7}, Equation (\ref{eq:closure})}
{Corollary~\ref{Corollary::7}(\ref{eq:closure})}
yields
$$\C{X}{P}{({ s}-1)E}* \C{X}{P}{{ s}E}^{\perp} =
\C{X}{P}{D_P-E+K} = \C{X}{P}{E}^{\bot}.
$$
\end{proof}

\subsection{A conjecture}
In \cite{wieschebrink:2010} Wieschebrink asserts, without proving it
that, with high probability, the square of a low codimensional subcode
$C$ of a $\mathrm{GRS}_k(\mathbf a, \mathbf b)$ is a GRS code.
More precisely, in general,
$C^{(2)} = \mathrm{GRS}_{2k-1}(\mathbf a, \mathbf b*\mathbf
b)$. Wieschebrink uses this observation to break Berger and Loidreau's
proposal \cite{berger:2005}.
It is natural to ask whether this property extends to low
  codimensional subcodes of AG codes. Some experimental results
  encourage us to establish the following conjecture.

\begin{conjecture}\label{conj:squares}
  Let $C$ be a uniformly random subcode of $\C{X}{P}{E}$ of
  dimension $\ell$ such that
$$\begin{array}{ccc}
2g+1\leq \deg(E) \leq \frac{n-1}{2} & \hbox{ and }&
2k+1-g \leq \binom{\ell+1}{2}
\end{array}$$
where $k=\deg(E) + 1-g$ is the dimension of $\C{X}{P}{E}$. Then,
the probability that $C^{(2)}$ is different from $\C{X}{P}{2E}$ tends to $0$ when $k$ tends to infinity.
\end{conjecture}

We give a proof along the lines of \cite[Remark 5]{marquez:2013} for
the special case of subcodes of GRS codes. Note that the case of GRS
codes is a bit different in terms of the probabilistic model since
for a GRS code, for the length to tend to infinity, the size of the alphabet
needs to tend to infinity too. 
{Evidences for this conjecture are discussed further below.}

% \sout{Our experimental results are in good agreement
% with this conjecture (see Table \ref{Table::3}).}
% \sout{See also} {\cite{mirandola:2012}.
% \sout{
% The following
% corollary is central to our attack against subcodes of AG codes.}

\begin{corollary}
  \label{Corollary::15}Assume that Conjecture \ref{conj:squares}
  holds.  Let $2g+1 \leq \deg(E) \leq \frac{n-2}{2}$.  Let $k \eqdef
  \deg(E) + 1 -g$ be the dimension of $\C{X}{P}{E}$, such that $2k
  +1-g\leq {\ell+1 \choose 2}$ for some $\ell$. Then, the equality
  $\overline{C}^{2} = \C{X}{P}{E}$ holds for uniformly random $\ell$-dimensional
  subcodes $C$ of $\C{X}{P}{E}$ with a probability tending to $1$ when
  $k$ tends to infinity.
\end{corollary}

 \begin{proof}Suppose  Conjecture \ref{conj:squares} holds.
Let $C$ be an $l$-dimensional subcode of $\C{X}{P}{E}$.
By Conjecture \ref{conj:squares}, we would have that $C^{(2)} = \C{X}{P}{E}^{(2)}$ with high probability.
Moreover $\C{X}{P}{E}$ is $2$-closed by Proposition \ref{Proposition::15}. Thus, applying Lemma \ref{Lemma::12}, we conclude that $\overline{C}^{2} = \C{X}{P}{E}$ which completes the proof.
 \end{proof}

{
\subsubsection*{Experiments around this conjecture}
To test the validity of the conjecture, we performed experiments as
follows:
\begin{itemize}
\item Generate a random smooth irreducible plane curve over $\F_q$
  using {\sc Magma} command {\tt RandomPlaneCurve};
\item Choose a divisor on the curve by selecting some random rational points and 
  places of higher degree and sum them up;
\item Compute the corresponding code $C$;
\item Choose a random subcode $C'$ of dimension $\ell$ such that $\ell$
  is the least integer satisfying
  ${\ell +1 \choose 2} \geq \dim C^{(2)}$;
\item Compare $C^{(2)}$ and $C'^{(2)}$.
\end{itemize}

\begin{remark}
  Note that in terms of the dimension $\ell$ of the subcode, we tested only
  the critical case i.e. the minimal value of $\ell$. Clearly cases with
  larger $\ell$'s can only be more successful.
\end{remark}

Such a test has been performed on $100$ random curves over fields $\F_q$
with $2 < q < 200$. These curve had genus $6 < g < 36$.
For each curve, and for any $2g+2 \leq m \leq \frac{n+g-2}{2}$ we chose
a divisor $G$ of degree $m$ and tested $10000$ random subcodes.
Among these $10000$ tests the codes $C'^{(2)}$ fails to equal $C^{(2)}$
at most $0.5 \%$ of the times. Moreover, for more than $90\%$ of pairs (curve, divisor), the failure rate is $0\%$.
}

%%% Local Variables: 
%%% mode: latex
%%% TeX-master: "Long-Version"
%%% End: 

\section{Decoding algorithms of algebraic geometry codes}
\label{Section4}
\subsection{Error-correcting pairs}

The notion of \emph{error-correcting pair} (ECP) for a linear code was introduced by Pellikaan \cite{pellikaan:1988,pellikaan:1992} and independently by K\"otter \cite{koetter:1992}.

Generally, given a positive integer $t$, a $t$--ECP for a linear code $C \subseteq \mathbb F_q^n$ is a pair of linear codes $(A,B)$ in $\mathbb F_q^n$ satisfying
$A*B\subseteq C^{\perp}$ together with several inequalities relating $t$ and the dimensions and (dual) minimum distances of $A, B, C$.
In a formal manner:

 \begin{definition}
\label{ECP}
 Let $C$ be a linear code in $\Fq^n$. A pair $(A, B)$ of linear codes over $\Fq^n$
 is called a \emph{t-error correcting pair} (ECP) for $C$ if the following conditions hold:
%\vspace{-.2cm}
\begin{multicols}{2}
\begin{enumerate}[(E.$1$)]
  \setlength{\itemsep}{-0.2mm}
 \item\label{it:E1} $(A*B) \subseteq C^{\bot}$,
 \item\label{it:E2} $k(A) > t$,
 \item\label{it:E3} $d(B^{\perp}) > t$,
 \item\label{it:E4} $d(A) + d(C) > n$.
 \end{enumerate}
\end{multicols}
 \end{definition}

It is shown that a linear code in $\Fq^n$ with a $t$-error correcting pair has a decoding algorithm which corrects up to $t$ errors with complexity $O \left(n^3\right)$.
ECPs provide a unifying point of view for several classical bounded distance decoding algorithms for algebraic and AG codes.
See \cite{marquez:2012b} for further details.

\begin{theorem}[{\!\!\cite[Theorem 3.3]{pellikaan:1992}}]
\label{Theorem::ECP}
Let $\X$ be a curve of genus $g$ and $E$ be a divisor on $\X$ such
that $n>\deg(E) > 3g - 1$.  Let $d^* \eqdef \deg (E) +2-2g$ be the
Goppa designed distance of $\C{X}{P}{E}^\bot$ and $t \eqdef
\left\lfloor {(d^*-1-g)}/{2}\right\rfloor$.  Let $F$ be any divisor
on $\X$ with disjoint support with $\DP$ and $\deg(F) = t+g$.
Then, the pair of codes defined by
$$\begin{array}{ccc}
A = \C{X}{P}{F} & \hbox{ and }& B = \C{X}{P}{E-F}
\end{array}$$
is a $t$-ECP for $\C{X}{P}{E}^{\bot}$.
Under the above conditions, such a divisor $F$ always exists.
\end{theorem}

\begin{corollary}
\label{Corollary::ECP}
Under the conditions stated above and assuming that $\deg(E)\leq n-3$
and $t\geq 1$. Then,
$$
A = {(B* C)}^{\perp}.
$$
\end{corollary}

\begin{proof}
Notice that $C= \C{X}{P}{E}^{\perp}$ and
\begin{eqnarray*}
\deg(E-F)  & = & \deg(E) - \deg(F) = \deg(E) - (t+g) = \deg(E) - 2t + t - g \\
& \geq  & \deg(E) - d^* +1 +t = 2g+t-1\geq 2g
\end{eqnarray*}
We now apply Proposition \ref{Proposition::8-9} to obtain the desired result.
\end{proof}

\begin{remark}\label{rem:fundamental}
%Let $t= \left\lfloor \frac{d^*-1-g}{2}\right\rfloor$.
  From a cryptanalytic point of view, the above corollary asserts
  that, it is sufficient to know the codes $\C{X}{P}{E}$ and $\C{X}{P}{E-F}$
  in order to get a $t$-ECP for $\C{X}{P}{E}$.
  Roughly speaking: {\em if you know $A$ and $C$, then you know $B$}.
\end{remark}

\subsection{Error-correcting arrays}

The notion of {\em majority voting of unknown syndromes} was initiated by Feng-Rao \cite{feng:1993}
for AG codes and by Feng-Tzeng \cite{feng:1994a} for cyclic codes. Duursma in \cite{duursma:1993a,duursma:1993b} treated it as {\em majority coset decoding}.

The philosophy of these algorithms can roughly be summarized as follows.
Suppose we have a code $C_1$ for which we need a decoding algorithm, and a
subcode $C_2$ for which we have a decoding algorithm.
{\em Coset decoding} is an algorithm which has as input a word
$\mathbf y_1$ such that  $\mathbf y_1\in \mathbf e +C_1$, and as output $\mathbf y_2$ such that $\mathbf y_2\in \mathbf e +C_2$.
In the present article, the code $C_2$ will always be $\{0\}$ for which there
exists an obvious decoding algorithm.

% Let $C_1=\C{X}{P}{mP}^{\perp}$ be a nonzero AG code of length $n$ and designed minimum distance $d_1^* = m-2g+1$ defined on a curve $\mathcal X$ of genus $g$ where $P$ is an $\mathbb F_q$-rational point of $\mathcal X$ with disjoint support from $\mathcal D_{\mathcal P}$. Consider its subcode $C_2=\C{X}{P}{(m+g)P}^{\perp}$ which has designed minimum distance $d_2=d_1^*+g$. Notice that, by Theorem~\ref{Theorem::ECP}, $C_2$ has a $t$-error correcting pair, where
% $$t=\left\lfloor \frac{d_2^*-1-g}{2} \right\rfloor = \left\lfloor \frac{d_1^*-1}{2} \right\rfloor .$$
% Thus, the idea of Feng-Rao is to apply \emph{coset decoding} to the code $C_1$ using its subcode $C_2$ from which we know an efficient decoding algorithm.
% The coset decoding is done in several steps, at each iteration, the dimension of the subcode
% drops by one.
% The whole setup is also used by Feng-Rao in another paper \cite{feng:1994b} to get an improved lower bound on the minimum distance of AG codes.

These algorithms have a purely linear algebraic description using the
notion of {\em error correcting arrays} (ECA) \cite{pellikaan:1993,
  lausten:1993, kirfel:1995} or that of {\em well behaving sequences}
\cite{geil:2013}. Take notice that error correcting arrays deal with
spaces of functions. In particular, in the case of AG codes, arrays
consist in infinite collections of Riemann Roch spaces, while well
behaving sequences are defined directly from error correcting codes
without involving any other external data.

In this article we chose to adopt a slightly different point of view
mixing the concepts of \cite{pellikaan:1993} and \cite{geil:2013}. For
that purpose we introduce the notion of {\em array of codes} which is
very similar to the notion of error correcting array but now defined
only with codes, without involving the function fields and Riemann
Roch spaces.  Such array of codes is strongly related to an error
correcting array in the sense of \cite{pellikaan:1993, lausten:1993,
  kirfel:1995}. Moreover, since we work only with codes, we also use
the notion of {\em well behaving pair} which is necessary for the
definition of the designed distance and for the decoding algorithm.
The choice of this mixed point of view is motivated by two facts.
First, from the cryptanalytic point of view, it is interesting
to show how to design a decoding algorithm from the single
data of a generator matrix of a code. On the other hand, the operation
we perform on codes arise from natural operations on Riemann Roch spaces.
For this reason, even if we do not directly compute Riemann Roch spaces
the language of arrays seemed more convenient than that of
well behaving sequences to describe and explain our calculations. 

% This is indeed interesting in terms of cryptanalysis: we show in the
% sequel how to design a decoding algorithm from the single data of a
% generator matrix of a given AG code.

\begin{definition}
\label{Definition:Array}
An \emph{array of codes} is a triple $(\mathcal A, \mathcal B,
\mathcal C)$ of sequences of linear codes
$\mathcal A = {(A_i)}_{1\leq i \leq n}$, $\mathcal B = {(B_i)}_{1 \leq i \leq n}$
and $\mathcal C = {(C_i)}_{1 \leq i \leq n}$ satisfying the following conditions
for all $i \in \{1, \ldots, n\}$,
\begin{enumerate}[({A}.1)]
\setlength{\itemsep}{-.2mm}
\item\label{item:array1} $\begin{array}{cccc}\dim(A_i) = i, & \dim(B_i) = i & \hbox{ and }&\dim(C_i) = n-i\end{array}$;
\item\label{item:array2} $\begin{array}{cccc}A_i \subseteq A_{i+1},& B_i \subseteq B_{i+1}& \hbox{ and }& C_i\supseteq C_{i+1}\end{array}$;
\item\label{item:array3} for all $r \in \{1, \ldots , n\}$,
there exists a pair $(i,j)$
such that $A_i * B_j \subseteq C_r^{\bot}$ and
$A_i * B_j \nsubseteq C_{r-1}^{\bot}$.
% \item For all $1 \leq i \leq u $ and $1 \leq j \leq v$ there exists an integer $r$ with $w \leq r \leq l$ such that $A_i * B_j \subseteq C_r^{\bot}$. We define
% $\hat{r}(i,j) = \min \left\{ r \mid A_i*B_j \subseteq C_r^{\bot}\right\}$.
% \item If $\hat{r}(i,j)$ is defined and $w< \hat{r}(i,j)$,
% then $\hat{r}(i,j)$ is strictly increasing in both arguments, i.e. $\hat{r}(i-1,j)< \hat{r}(i,j)$ for all $i\geq 1$ and $\hat{r}(i, j-1)< \hat{r}(i,j)$ for all $j\geq 1$.

% \item If $\mathbf a \in A_i \setminus A_{i-1}$ and $\mathbf b \in B_{j}\setminus B_{j-1}$, then
% $\mathbf a * \mathbf b \in C_r^{\bot}\setminus C_{r-1}^{\bot}$ with  $r=\hat{r}(i,j)>w$.
\end{enumerate}
\end{definition}

In addition we introduce the function $
\hat{r}:  \{1, \ldots , n\}^2  \rightarrow \{1, \ldots ,n\}$ defined as follows.

\begin{definition}
  Let $(\mathcal A, \mathcal B, \mathcal C)$ be an error correcting array.
  The function $\hat{r}$ is defined as
  $$
  \hat{r}(i,j) \eqdef \min \left\{ r \in \{1, \ldots, n\} ~\Big|~
  A_i * B_j \subseteq C_r^{\bot}\right\}.
  $$
% \min_{r \in \{1, \ldots ,  n\}} \left\{A_i * B_j \subseteq C_r^{\bot}
% {\color{blue}\mid 1\leq i, j \leq n}
% \right\}.
\end{definition}

\begin{remark}
  Condition (A.\ref{item:array3}) of arrays of codes asserts that
  $\hat r$ is surjective.
\end{remark}

\begin{remark}
Note that $\hat{r}$ is increasing in both arguments but not necessarily
strictly increasing. In particular if for some pair $(i,j) \neq (n,n)$
we have $\hat{r}(i,j) = n$ then, it is clear that $\hat{r} (k,l) = n$
for all $k\geq i$ and $l \geq j$.  
\end{remark}

% Notice that $\hat{r}(i,j)$ is a partial function, that is to say, it is only defined for those pairs $(i,j)$ such that
% there exists an $r$ with $w \leq r \leq l$ and  $A_i * B_j \subseteq C_r^{\bot}$.
% If $\hat{r}(i,j)$ is defined, then $\hat{r}(i',j')$ is also defined for all $i'\leq i$ and $j'\leq j$.

\begin{definition}
  A pair $(i,j) \in \{1, \ldots ,n\}^2$ is said to be {\em well behaving}
  (WB in short) if
  $$
  \forall (i',j') \ {\rm such \ that}\ i\leq i', j\leq j'
  {\rm \ and\ } (i,j)\neq (i',j'),\ 
  \hat r(i',j') < \hat r(i,j).
  $$
  For all $r \in \{1, \ldots, n\}$, set
  \begin{itemize}
    \setlength{\itemsep}{-.2mm}
    \item $\hat{n}_r \eqdef \big| \left\{ (i,j) \mid 1\leq i,\ j \leq n, \ 
      (i,j)\ {\rm is\ WB\ and\ } \hat{r}(i,j) = r+1 \right\}\big|;$
    %\item $\hat n_r  \eqdef |\hat N_r|; $
    \item $\hat{d}_r  \eqdef \min \left\{\hat{n}_{r'} \mid r\leq r' \leq n
    \right\}.$
  \end{itemize}

\end{definition}
We have the following result:

\begin{theorem}
For any array of codes $(\mathcal A, \mathcal B, \mathcal C)$ we have
$$\hat{d}_r \leq d(C_r)\hbox{, for all } 1\leq r \leq n.$$
\end{theorem}

\begin{proof}
  The proof is very similar to that of
  \cite[Theorem 2.5]{kirfel:1995}.
\end{proof}

\begin{definition}
\label{ECA}
An array of codes $(\mathcal A, \mathcal B, \mathcal C)$ in $\F_q^n$
%  enumerated by
% $$
% \begin{array}{cccc}
% \mathcal A = (A_i)_{1\leq i \leq u}, &
% \mathcal B = (B_j)_{1\leq j \leq v} & \hbox{and} &
% \mathcal C = (C_r)_{w\leq r \leq l}
% \end{array}
% $$
is said to be $t$-{\em error correcting} for a code $C$ in $\F_q^n$ if
there exists an $s$ with $1\leq s \leq n$ such that $C = C_s$
and $ t\leq \frac{\hat{d}_s-1}{2}\cdot$
% \item and one of the following conditions is verified:
% \begin{itemize}
% \item $C_l = 0$
% \item There exists $i$ and $j$ with $1 \leq i \leq u$ and $1 \leq j \leq v$ such that \\
% $(A_i, B_j)$ is a $t$-ECP for $C_r$ with $s\leq r\leq l$.
% \end{itemize}
\end{definition}

% \begin{remark}\todo{remark added}If $(A_i, B_j)$ is a $t$-ECP for $C_r$, then $A_i*B_j\subseteq C_r^\perp$.
% Hence $\hat{r}(i,j)\leq r$.
% \end{remark}

\begin{theorem}
A linear code $C$ in $\mathbb F_q^n$ with a $t$-ECA has a decoding algorithm
which corrects up to $t$ errors with complexity $\mathcal O(n^3)$.
\end{theorem}

\begin{proof}
See \cite{feng:1993} and \cite[Theorem 2.9]{kirfel:1995}.
% The proof is in \cite{feng:1993} and \cite[Theorem 2.9]{kirfel:1995}. The reason is very simple:
% suppose we get the received word $\mathbf y = \mathbf e + \mathbf c$ with $\mathbf c\in C$
% but we do not know a decoding algorithm for $C$.
% However, we have $\mathbf y_r = \mathbf e + \mathbf c_r$ with $\mathbf c_r \in C_r \subseteq C$ for all $s \leq r \leq l$, by a majority vote of unknown syndromes.
% \begin{itemize}
% \item In the first case, if $C_l = 0$, then we are done since $\mathbf y_l = \mathbf e$.
% \item In the second case, if $C_r$ has a $t$-ECP, then we apply the corresponding decoding algorithm to $\mathbf y_r$ with the pair
% $(A_i, B_j)$ to obtain the error vector $\mathbf e$.
% \end{itemize}
\end{proof}

\begin{remark}
  In the literature, for instance in \cite{pellikaan:1993} or
  \cite[Remark 2.10]{kirfel:1995}, the definition of a $t$--error
  correcting array for a code $C$ is more general than that of this
  article.  Indeed, usually, an ECA for a code $C$ is associated to a
  sequence of codes ${(C_i)}_{\ell \leq i \leq u}$ where $C = C_r$ for
  some $r\in [\ell,u]$ and the code $C_\ell$ is either zero or has a
  $t$--ECP.  This condition is sufficient to get a decoding
  algorithm. Indeed, if we receive the word $\mathbf y = \mathbf e +
  \mathbf c$ with $\mathbf c\in C$ and $\mathbf e$ has weight less than or
  equal to $t$, then, by coset decoding, we obtain
  a vector $\mathbf y_\ell$ such that $\mathbf y_\ell = \mathbf e +
  \mathbf c_\ell$ for some $\mathbf c_\ell \in C_\ell$. Next,
  either $C_\ell = \{0\}$ and we get directly $\mathbf e$ or $C_\ell$
  has a $t$--ECP which we can use to obtain $\mathbf e$.

  We chose to avoid such a general definition since it is useless for
  our purpose.
\end{remark}

\subsubsection*{The array of interest in this article}\label{ss:array_of_interest}
Let $E$ be a divisor on $\X$ and $P$ be a rational point of $\X$.
Let $(\alpha_i)_{i \in \mathbb N}$ be the non gap sequence at $P$ and $(\beta_i)_{i \in \mathbb N}$ the $E$--non gap sequence at $P$ 
(see Section~\ref{Weierstrass-gaps} for further details on Weierstrass gaps).
We introduce the {\bf finite} sequences $(\hat \alpha_i)_{1 \leq i \leq n}$
and $(\hat \beta_i)_{1 \leq i \leq n}$ defined as follows:
\begin{itemize}
\setlength{\itemsep}{-.02mm}
\item for all $i \in \{1, \ldots, n\}$, $\hat \alpha_i$ 
  is the least integer such that $\dim \C{X}{P}{\hat \alpha_i P} = i$;
\item for all $i \in \{1, \ldots, n\}$, $\hat \beta_i$ 
  is the least integer such that $\dim \C{X}{P}{E + \hat \beta_i P} = i$.
\end{itemize}

\begin{remark}
For indexes $i$ such that the evaluation map is injective, i.e. such that
$L(\alpha_i P - D) = 0$ (resp. $L (E + \beta_i P - D) = 0$), 
we have
$\hat \alpha_i = \alpha_i$ (resp. $\hat \beta_i = \beta_i$) while for larger
dimensions, we need to consider the contribution of $(-D)$--non-gaps at $P$
(resp. $(E-D)$--non-gaps). % Nevertheless we always have
% $$
% \forall  i \in \{1, \ldots,  n\}, \quad \hat \alpha_i \geq \alpha_i,\quad {\rm
% and}\quad \hat \beta_i \geq \beta_i.
% $$  
\end{remark}

We define the triple $(\mathcal A, \mathcal B, \mathcal C)$
of sequences of linear codes in $\F_q^n$ as:
\begin{itemize}
\setlength{\itemsep}{-.02mm}
\item $\forall i \in \{1, \ldots, n\},\ \ A_i \eqdef \C{X}{P}{\hat \alpha_i P}$;
\item $\forall j \in \{1, \ldots, n\},\ \ B_j \eqdef \C{X}{P}{E+\hat \beta_jP}$;
\item $\forall r \in \{1, \ldots, n\},\ \ C_r \eqdef \C{X}{P}{E+\hat\beta_rP}^{\bot} = B_r^{\bot}$.
\end{itemize}

\begin{proposition}
\label{Main-Part1}
The above defined triple $(\mathcal A, \mathcal B, \mathcal C)$
is an array of codes.
\end{proposition}

\begin{proof}
Conditions (A.\ref{item:array1}) and (A.\ref{item:array2}) are direct consequences
of the definition of the $\hat \alpha_i$ and $\hat \beta_j$'s.
Moreover, it is easy to see that $A_1$ is the repetition code i.e.
the code spanned by $(1, \ldots, 1)$. Hence for all $1 \leq i \leq n$, we have
$A_1 * B_{n-i} = B_{n-i} = C_i^{\bot}$ which gives (A.\ref{item:array3}).
\end{proof}

\begin{theorem}
  Assume that $\deg (E) < n$.
  Let $d^* = \deg (E) + 2 - 2g$ be the Goppa
  designed distance of $\C{X}{P}{E}^{\bot}$ and set
  $t = \lfloor (d^* - 1)/2 \rfloor$. Then, the above described triple
  $(\mathcal A, \mathcal B, \mathcal C)$ is a $t$--ECA for 
  $\C{X}{P}{E}^{\bot}$.
\end{theorem}

\begin{proof}
  Clearly, $\C{X}{P}{E}^{\bot}$ is an element of the sequence
  ${(C_i)}_{1\leq i \leq n}$. Let $k$ be the dimension of
  $\C{X}{P}{E}^{\bot}$.  By definition, we have $\C{X}{P}{E}^{\bot} =
  C_{n-k}$ and $\hat \beta_{n-k} = 0$.  The only thing we need to prove is
  that $\hat d_{n-k} \geq d^*$. This can be proved in a very similar
  fashion as \cite[Corollary 3.9]{kirfel:1995}
  which is a direct consequence of \cite[Theorem 3.8]{kirfel:1995}. 
  The cited proof involves the set of pairs $(i,j)$ such that
  \begin{equation}\label{eq:sum_for_non_gaps}
    \alpha_i + \beta_j = \beta_{n-k}.
  \end{equation}
  If we prove that for any such pair $(i,j)$ we have
  $\alpha_i = \hat \alpha_i$ and $\beta_j = \hat \beta_j$, then
  the proof of  \cite[Theorem 3.8]{kirfel:1995} will apply
  mutatis mutandis in our setting.

  Let $(i,j)$ be a pair of positive integers satisfying
  (\ref{eq:sum_for_non_gaps}).
  Since we always have $\alpha_i \geq 0$,
 %\todo{there exists instead of we have} 
  there exists $\beta_j \leq \beta_{n-k} = 0$. 
  But, the assumption $\deg (E) < n$,
  entails that $\deg (E + \beta_j P - D) < 0$
  and hence $L(E - \beta_j P - D) = \{0\}$.
  Consequently, $\beta_j = \hat \beta_j$.
  On the other hand we always have $\beta_j \geq - \deg (E)$ and hence
  $\alpha_i \leq \deg (E) < n$ which entails $L(\alpha_iP - D) = \{0\}$
  and gives that $\hat \alpha_i = \alpha_i$.
\end{proof}

\section{The $P$-filtrations}
\label{Section5}
Let $P$ be one of the points of the $n$-tuple $\P$. Let $E$ be a
divisor on $\X$ that has disjoint support from $\P$. Let $\P'$ be the
$(n-1)$--tuple obtained from $\P$ by deleting $P$. In the present
section we give an efficient
way to obtain a $t_P$--ECP and a $t_A$--ECA for the code
$\C{X}{P'}{E}^{\bot}$ with
$$\begin{array}{cccc}
d^* = \deg(E) -2g+2, &
t_P = \left\lfloor \frac{d^*-g-1}{2} \right\rfloor & \hbox{ and }&
t_A = \left\lfloor \frac{d^*-1}{2}\right\rfloor
\end{array}.$$
By this manner, given a received word $\mathbf y = \mathbf c + \mathbf
e$ where $\mathbf c \in \C{X}{P}{E}^{\bot}$ and $\mathbf e$ has weight
$\leq t_P$ (resp. $\leq t_A$), one can proceed to decoding as follows.
Thanks to our ECP (resp. ECA), one decodes the word $\mathbf y$
punctured at position $P$, which yields $\mathbf c$ punctured at this
position. Then retrieving $\mathbf c$ consists only in correcting an
erasure at position $P$.

\begin{remark}
  A method to deduce directly an ECP for the non punctured code
  is described in Appendix~\ref{Degenerate}.
\end{remark}

From the knowledge of a generator matrix of $\C{X}{P}{E}^{\bot}$ we
aim at computing the sequences of codes ${(A_i)}_{1 \leq i \leq n}$
and ${(B_j)}_{1 \leq j \leq n}$ introduced in
Section~\ref{ss:array_of_interest}.  For the computational aspects it is
convenient to introduce another pair of sequences very similar to
${(A_i)}_i$ and ${(B_i)}_i$ but with different indexes.  Namely, we
introduce the sequences ${(U_i)}_{i \in \Z}$ and ${(V_i)}_{i \in \Z}$
of codes of length $n-1$ defined as
$$
\forall i \in \Z,\quad U_i \eqdef \C{X}{P'}{iP}
\qquad {\rm and} \qquad V_i \eqdef \C{X}{P'}{E+iP}.
$$
These sequences are related to the $A_i$'s and $B_j$'s by the relation:
$$
\forall i \in \{1, \ldots, n\}, \quad A_i = U_{\hat \alpha_i} 
\qquad {\rm and} \qquad B_i = V_{\hat \beta_i}.
$$
The reader will observe that, despite the introduction of new notation,
this other way of indexing the sequences is more convenient for
explicit computation and for understanding the behaviour of these
codes with respect to the Schur product.

\subsection{Structure of the sequences ${(U_i)}_i$ and
${(V_i)}_{i}$}
First one observes that for all $i< 0$, $U_i = \{0\}$.
On the other hand for all $i \geq n+(2g)-2$ one can prove that
$U_i = \F_q^{n-1}$. Indeed, set
$$\DPP \eqdef \sum_{P \in \P'} P.$$ 
Then, for all integer $i$, the evaluation map
$\ev_{\P'}$ induces an isomorphism between $U_i$ and $L(iP)/L(iP-\DPP)$
and Riemann Roch theorem asserts that $\dim L((n+2g-2)P) = n-1+g$ and
$\dim L ((n+2g-2)P - \DPP) = g$ which proves that $\dim U_{n+2g-2} = {n-1}$
and hence $U_{n+2g-2} = \F_q^{n-1}$.

Next, one can split the sequence ${(U_i)}_{0 \leq i \leq n+2g-2}$ in three parts
represented by the three following diagrams.
A first part ${(U_i)}_{0 \leq i \leq 2g-2}$ in which some consecutive terms may be equal because of 
gaps at $P$:
$$
\xymatrix{
\relax \{0\}  \ar@{^{(}->}[r] 
& L(0)  \ar@{^{(}->}[r] \ar[d]^{\rotatebox{90}{$\sim$}}_{\ev_{\P'}} &
L(P) \ar@{^{(}->}[r] \ar[d]^{\rotatebox{90}{$\sim$}} & \ \ \cdots \ \
 \ar@{^{(}->}[r]
& L((2g-1)P) \ar@{^{(}->}[r]_-{\neq} \ar[d]^{\rotatebox{90}{$\sim$}}&
L(2gP) \ar@{^{(}->}[r]_-{\neq} \ar[d]^{\rotatebox{90}{$\sim$}} & \ \ \cdots \\
\{0\} \ar@{^{(}->}[r] & U_0 \ar@{^{(}->}[r] & U_1 \ar@{^{(}->}[r] &
\ \ \cdots \ \  \ar@{^{(}->}[r] & U_{2g-1}  \ar@{^{(}->}[r]_-{\neq} & U_{2g}
\ar@{^{(}->}[r]_-{\neq} & \ \ \cdots
} 
$$
A second part ${(U_i)}_{2g-1 \leq i \leq n-2}$ which is regular, i.e. any term
has codimension 1 in the next one:
$$
\xymatrix{ \cdots \ \ \ar@{^{(}->}[r] &
\relax L((2g-1)P) \ar@{^{(}->}[r]_-{\neq} \ar[d]^{\rotatebox{90}{$\sim$}}_{\ev_{\P'}}
& \relax L(2gP) \ar@{^{(}->}[r]_-{\neq}
\ar[d]^{\rotatebox{90}{$\sim$}} & \ \ \cdots \ \
\ar@{^{(}->}[r]_-{\neq} & L((n-2)P) \ar@{^{(}->}[r]_-{\neq}
\ar[d]^{\rotatebox{90}{$\sim$}} & \ \ \cdots \\
\cdots\ \ \ar@{^{(}->}[r] & U_{2g-1}  \ar@{^{(}->}[r]_-{\neq} & 
U_{2g}  \ar@{^{(}->}[r]_-{\neq} & \ \ \cdots \ \  \ar@{^{(}->}[r]_-{\neq} & 
U_{n-2}  \ar@{^{(}->}[r]_-{\neq} & \ \ \cdots
}
$$
Finally, a third part ${(U_i)}_{n-1 \leq i \leq n+2g-2}$ where the map $\ev_{\P'}$
stops to be injective, but remains surjective. In this range, consecutive terms
may be equal: it happens at every $(-\DPP)$--non-gap:
$$
\xymatrix{
\relax
 \cdots \ \ \ar@{^{(}->}[r] & L((n-1)P) \ar@{^{(}->}[r] \ar@{->>}[d]_{\ev_{\P'}}&
\ \ \cdots \ \ 
\ar@{^{(}->}[r]  & L((n+2g-3)P)  \ar@{^{(}->}[r] \ar@{->>}[d] &
L((n+2g-2)P)  \ar@{->>}[d]  \\
\cdots \ \ \ar@{^{(}->}[r] & U_{n-1}  \ar@{^{(}->}[r] & 
\ \ \cdots \ \  \ar@{^{(}->}[r] & U_{n+2g-3}  \ar@{^{(}->}[r] & \F_q^{n-1}
}
$$

In the very same manner, for any $i < - \deg (E)$, $V_i = \{0\}$
and for any $i\geq 2g-2+n-\deg(E)$.
Next, the sequence splits in three parts:
\begin{itemize}
    \setlength{\itemsep}{-.5mm}
\item A first part ${(V_i)}$ where ${-\deg(E)+1 \leq i \leq -\deg(E)+2g-1}$
in which consecutive terms are equal at each $E$--gap at $P$;
\item a second part ${(V_i)}$ where ${-\deg(E)+2g \leq i \leq -\deg (E) + n-2}$
which is regular, i.e. any two consecutive terms are distinct;
\item and a third part ${(V_i)}$ where
${-\deg (E) + n-1 \leq i \leq -\deg (E) + n +2g-2}$
in which consecutive terms equal at each $(E-\DPP)$--non--gap.
\end{itemize}

\subsection{Effective computations}

Here we explain how to compute the terms of this sequence only by performing
Schur products and solving linear systems. Notice that very similar
methods have been used by Khuri--Makdisi to perform effective
computations on Jacobians of curves \cite{khuri2001}.

\subsubsection{Which elements of the sequence do we know on the beginning?}
\label{ss:which}
From a generator matrix of $\C{X}{P}{E}^{\bot}$, one can compute
$\C{X}{P}{E}$. Then, $V_0$ and $V_{-1}$ are
obtained from the code $\C{X}{P}{E}$ respectively by puncturing and shortening
at the position $P$. All these operation boil down to
Gaussian elimination.

\subsubsection{Computing terms of ${(V_i)}_i$ from other terms
of ${(V_i)}_i$}

The following statement explains how to compute $V_{-i-1}$ (resp. $V_{i+1}$)
from the knowledge of the codes $V_{-i}$, $V_{-i+1}$ (resp. $V_{i}$, $V_{i-1}$).

\begin{proposition}
  \label{prop:comput_Vi}
  Let $i \geq 1$, then we have
  \begin{enumerate}[(i)]
    \setlength{\itemsep}{-.5mm}

\item\label{item:pb1} if $\deg (E) - \frac{n-4}{2} \leq i \leq
    \deg (E) - 2g +1$, then 
    
 % {\color{blue}
 %    \textbf{Explanation - Must be deleted in the final version:}
    
 %    By Proposition 7 we should have that:
 %    $$\deg(E+(-i+1)P) \geq 2g \Longrightarrow -i \geq 2g-\deg(E)-1 \Longrightarrow i \leq \deg(E) - 2g +1$$
 %    }
    \begin{equation}\label{Problem::1}
      V_{-i-1} = \left\{\mathbf z \in V_{-i} ~\big|~ \mathbf z *
    V_{-i+1} \subseteq V_{-i}^{(2)}\right\};
    \end{equation}

\item\label{item:pb2} if $\frac{n-4}{2}- \deg (E) \geq i \geq 2g+1-\deg(E)$,
    then
    \begin{equation}\label{eq:pb2}
    V_{i+1} = \left\{\mathbf z \in \F_q^{n-1} ~\big|~ \mathbf z *
    V_{i-1} \subseteq V_{i}^{(2)}\right\}.
    \end{equation}
  \end{enumerate}
\end{proposition}

\begin{proof}
Assume that $\deg (E) - \frac{n-4}{2} \leq i \leq
    \deg (E) - 2g - 1$.
By %\sout{Lemma \ref{Lemma::9}} {\color{red}
Proposition~\ref{Proposition::8-9}, the solution space of (\ref{Problem::1}) is equal to
$$V_{-i} \cap \left( V_{-i+1} * \left(V_{-i}^{(2)}\right)^{\bot}\right)^{\bot}.$$
Since $\deg(E-iP) \geq 2g+1$, then,
from Corollary~\ref{Corollary::7}, we have $V_{-i}^{(2)} =
\C{X}{P'}{2E-2iP}$. Moreover, since
$
\deg(2E-2iP) \leq (n-1) -3$,
from Proposition \ref{Proposition::8-9} we conclude that
$$\left(V_{-i+1}* \left(V_{-i}^{(2)}\right)^{\bot}\right)^{\bot} =
\C{X}{P'}{(2E-2iP) - (E-(i-1)P)} = V_{-i-1}.$$
This proves (\ref{item:pb1}).
The proof of (\ref{item:pb2}) is very similar.
\end{proof}

Therefore, we can define an algorithm for determining the code
$V_{-i}$ for $i\geq 1$ wich consists in $i$ repeated applications of
Proposition~\ref{prop:comput_Vi}. But we can do better by decreasing
the number of iterations and relaxing the parameters conditions using
the following generalization of Proposition~\ref{prop:comput_Vi}
whose proof is very similar.

\begin{proposition}
  \label{prop:Comput_Vi_bis}
  Let $a \leq b \leq c \leq d$ be integers such that $a+d = b+c$.
  If $\deg (E) + b \geq 2g+1$ and $2 \deg (E) +b+c \leq n-4$ then,
  \begin{enumerate}[(i)]
    \setlength{\itemsep}{-.2mm}
  \item \label{it:compVi_bis}
    $
    V_a = \left\{ \mathbf z \in V_b ~\big|~ \mathbf z * V_d \subseteq V_b * V_c\right\}.
    $
  \item If moreover $\deg (E) + a \geq 2g$, then 
    $
    V_d = \left\{ \mathbf z \in \F_q ^{n-1} ~\big|~ \mathbf z * V_a \subseteq V_b * V_c\right\};
    $
  \end{enumerate}
\end{proposition}

\begin{remark}\label{rem:quick_comput}
  The previous statement permits for instance to compute $V_{-i}$
  as 
  $$
  V_{-i} = \left\{ \mathbf z \in V_{\lfloor \frac {-i-1} 2 \rfloor} ~\Big|~ 
  \mathbf z * V_0 \subseteq V_{\lfloor \frac {-i-1} 2 \rfloor} * 
  V_{\lfloor -\frac {-i+1} 2 \rfloor} \right\}.
  $$
  In the same spirit as the quick exponentiation algorithm, the recursive
  application of the above formula allows to compute $V_{-i}$ in
  $O(\log(i))$ iterations of Proposition~\ref{prop:Comput_Vi_bis}
  instead of $O(i)$ iterations of Proposition~\ref{prop:comput_Vi}.
\end{remark}

\subsubsection{Further computations}

In the same manner if we know some terms of the sequence ${(U_i)}_i$
we can compute other ones as follows.

\begin{proposition}
  \label{prop:Comp_Ui_from_Ui}
  Let $i,j$ be integers.
  \begin{enumerate}[(i)]
    \setlength{\itemsep}{-0.2mm}
  \item\label{it:Ui_from_Ui_1} If $i \geq 2g$ and $j \geq 2g+1$ then
    $U_i * U_j = U_{i+j}$.
  \item\label{it:Ui_from_Ui_2} Let $\ell$ be an integer such that
    $i+\ell = 2j$. If $i \geq 2g$ and $2g+1 \leq j \leq \frac{n-4} 2$,
    then
    $$
    U_\ell = \left\{ \mathbf z \in \F_q^{n-1} ~\big|~ \mathbf z * U_i
      \subseteq U_j^{(2)} \right\}.
    $$
  \end{enumerate}
\end{proposition}

It is also possible to compute some terms of one of the sequence
from the knowledge of terms of the other sequence:
\begin{proposition}
  \label{prop:Comp_Ui_from_Vi}
  Let $i, j, \ell$ such that { $i+j = \ell$, }
  $\deg (E) + \ell \leq n-4$.  Then,
  \begin{enumerate}[(i)]
    \setlength{\itemsep}{-0.2mm}
  \item\label{it:Ui_from_Vi} { If $\deg(E)+j \geq 2g$,
      then}
    $\ \ U_i = \left\{ \mathbf z \in \F_q^{n-1} ~\big|~ \mathbf z *
      V_j \subseteq V_\ell\right\}; $
\item\label{it:Vi_from_Ui} { If $j \geq 2g$, then}
  $\ \ V_i = \left\{ \mathbf z \in \F_q^{n-1} ~\big|~ \mathbf z * U_j
    \subseteq V_\ell\right\}.  $
  \end{enumerate}
\end{proposition}

Finally, notice that some terms of ${(V_i)}_i$ can be constructed using
the following statement.

\begin{proposition}
  \label{prop:Comp_Vi_from_U_iV_j}
  For all { $i \geq 2g$ and $j \geq 2g - \deg (E)$ with either $i > 2g$ or $j> 2g - \deg(E)$}, we have
  $V_{i+j} = U_i * V_j$.  
\end{proposition}

\subsubsection{Complexity}\label{ss:complexity}
The computation of one of the $U_i$'s or $V_i$'s using one of the previous
statement consists in computing a finite number of Schur products. The cost of
the computation of a Schur product is $O(n^4)$ (see for instance
\cite[Proposition 5]{CGGOT12}). 
A probabilistic shortcut allows to reduce the complexity of this computation
to $O(n^{3+\varepsilon})$ operation for $\varepsilon > 0$ arbitrarily small.
Indeed, given a code $C$ of length $n$ and dimension $k$, the computation
of $C^{(2)}$ consists in computing the ${k+1 \choose 2} = O(n^2)$
generators and then, to perform Gaussian elimination to deduce a basis
from this family of generators which costs $O(n^4)$.
However, by extracting $n+ \varepsilon$ elements chosen at random
from this set of generators, we get another generating set with a
large probability and this probabilistic trick reduces the complexity
to $O(n^{3 + \varepsilon})$.
A similar probabilistic shortcut is used in \cite{khuri2007}.

%%% Local Variables: 
%%% mode: latex
%%% TeX-master: "Long-Version"
%%% End: 

%\input{Section5}

\section{The Attack}
\label{Section6}

\subsection{The McEliece encryption scheme}
Let $\mathcal F$ be a family of linear codes with an efficient decoding algorithm. Every element of this family is represented by the triple $(C, \mathcal A_{C},t)$ where $\mathcal A_{C}$ denotes a decoding algorithm for $C\in \mathcal F$ which corrects up to $t$ errors. The McEliece scheme can be summarized as follows:
Alice applies an encoding mechanism to a message and adds enough errors to make it unintelligible. Then, Bob is the only person that knows an efficient decoding method (the secret key) to detect and correct those errors. That is:

\begin{itemize}
\item[]\textbf{Key generation:} Consider any element $(C, \mathcal A_{C},t)\in \mathcal F$. Let $G$ be a generator matrix of $C$. Then the \emph{public key} and the \emph{private key} are given respectively by
$$\begin{array}{ccc}
\mathcal K_{\mathrm{pub}} = (G,t) & \hbox{ and }&
\mathcal K_{\mathrm{secret}} = \mathcal A_{C}.
\end{array}$$

\item[]\textbf{Encryption:} The plaintext $\mathbf m$ is encrypted as
  $\mathbf y = \mathbf m G + \mathbf e$ where $\mathbf e$ is a random
  error vector of weight at most $t$.

\item[]\textbf{Decryption:} Using $\mathcal K_{\mathrm{secret}}$, the receiver obtains $\mathbf m$.
\end{itemize}

% Thus, code-based cryptography is based on the following one-way trapdoor function:
% \begin{itemize}
% \item It is easy and fast to encode a message using linear transformations, that is,  matrix multiplication.
% \item It is hard to decode random linear codes. Indeed, the general decoding problem was proven to be NP-complete in 1978 \cite{berlekamp:1978} for the Hamming metric.
% \item The trapdoor information is that there exists families of codes that have efficient decoding algorithms. Decoding is also believed to be hard for a quantum computer.
% \end{itemize}

\subsection{Context of the present article}

\label{sec:context}
In what follows, $\mathcal{X}$ denotes a smooth projective geometrically connected curve
over $\F_q$ of genus $g$,  $\P = (P_1, \ldots , P_n)$ denotes
an $n$-tuple of mutually distinct $\mathbb F_q$-rational points of $\mathcal X$, $\DP$ denotes the divisor $\DP \eqdef  P_1 + \cdots + P_n$ and $E$ denotes a divisor with disjoint support from that of $\DP$.

We assume that our public key is a generator matrix $\Gm$ of the public code $\C{X}{P}{E}^{\bot}$ and the largest number $t$ of errors introduced during the encryption step.
We take $t\leq\left\lfloor ({d^{*}-1})/{2}\right\rfloor$
where $d^{*}=\deg (E)-2g+2$ is called the \emph{designed minimum distance} of the public code $\C{X}{P}{E}^{\bot}$.
Thus,
$$
C_{pub} ~~:~~ \mathbf G \hbox{ a generator matrix of }
\C{X}{P}{E}^{\bot} \hbox{ and }t.
$$
Our attack consists in the computation either of an \emph{error-correcting
  pair} (ECP) or of an \emph{error-correcting array} (ECA) in order to
decode $\C{X}{P}{E}^{\bot}$. For this sake,
we distinguish two different cases:
$$\begin{array}{ccc}
t_P\leq \frac{d^*-g-1}{2} \hbox{ (i.e. related to ECP)} & \hbox{ and }&
t_A\leq \frac{d^*-1}{2} \hbox{ (i.e. related to ECA)}.
\end{array}$$

Take notice that, from the single knowledge of a generator matrix of $\C{X}{P}{E}$, one can compute $\deg(E)$ and the genus $g$ of $\X$ using the following statement.

\begin{proposition}
[\!{\cite[Proposition 18]{marquez:2013b}}]
\label{Genus}
If $2g+1\leq \deg(E) { < } \frac{n}{2}$. Let $k_1$ and
$k_2$ be the dimension of $\mathcal C = \C{X}{P}{E}$ and
$\mathcal C^{(2)}$, respectively. Then,
$$\begin{array}{ccc}
\deg(E) = k_2 - k_1 & \hbox{ and } & g= k_2 -2k_1 +1.
\end{array}$$
\end{proposition}

\subsection{In case $t\leq \frac{d^*-g-1}{2}$, i.e. computing an ECP}
\label{ss:attack_ECP}

In this section we describe how to attack the McEliece cryptosystem
based on AG codes when $t \leq \frac{d^*-g-1}{2}$.  If
$\frac{n}{2}-2\geq \deg(E)\geq 3g+t-2$, then the attack summarizes as
follows. The upper bound on $\deg (E)$ can be relaxed by applying techniques
from Section~\ref{Extending}. { On the other hand, if $\deg (E)$
is below the lower bound $3g+t-2$, one can still compute an ECA using the
techniques of Section~\ref{ss:attack_ECA}, which provides a more efficient
decoding algorithm.}

{
  \begin{remark}
    Note that for $t$ to be positive, the degree of $E$ should satisfy
    $$
    \deg (E) \geq 3g.
    $$
    Indeed, we have $t \leq \frac {d^*-g-1}2 = \frac{\deg (E) - 3g + 1}{2}$.
  \end{remark}
}

\begin{description}
\setlength{\itemsep}{-.2mm}
\item \textbf{Step 1.} Determine the values $g$ and $\deg(E)$ using Proposition \ref{Genus}.
\item \textbf{Step 2.} Compute $V_0 = \C{X}{P'}{E}$ and $V_{-1} = \C{X}{P'}{E-P}$
 by Gaussian elimination.
\item \textbf{Step 3.} Compute the code $V_{-t-g} =
  \C{X}{P'}{E-(t+g)P}$ using Proposition~\ref{prop:comput_Vi}(\ref{item:pb1})
  or Remark~\ref{rem:quick_comput}.
\item \textbf{Step 4.} Apply Corollary \ref{Corollary::ECP} to deduce an ECP for $\mathcal C_{\mathrm{pub}}$ (punctured at the position $P$).
\end{description}

\begin{remark}
  Remind that the above procedure provides an ECP for the code
  $\C{X}{P}{E}^{\bot}$ punctured  at one position.
  But as explained in Section~\ref{Section5},
  the decoding of  $\C{X}{P}{E}^{\bot}$ can be performed by first correcting
  errors on the punctured code and then correct an erasure on 
  $\C{X}{P}{E}^{\bot}$.
  A method to get directly an ECP for $\C{X}{P}{E}^{\bot}$ is presented
  in Appendix~\ref{Degenerate}.
\end{remark}

\paragraph{Complexity}
The costly part of the procedure is the calculation of $V_{-t-g}$.
If we proceed to $t+g$ iterations of
Proposition~\ref{prop:comput_Vi}(\ref{item:pb1}), then
from Section~\ref{ss:complexity}, the complexity is $O((t+g)n^4)$.
Using Remark~\ref{rem:quick_comput} the cost can be reduced to
$O(\log (t+g) n^4)$. Then, using the probabilistic shortcut explained in
Section~\ref{ss:complexity} we get a complexity in
$O(\log(t+g)n^{3 + \varepsilon})$.

% %%%%% BEGIN algorithm 2
% \begin{algorithm2e}[!h]
% \KwData{Generator matrices for the codes $V_0$ and $V_1$}
% \KwResult{A generator matrix for the code $V_{t+g}$}

% Let $b= \left\lfloor \log_2 n\right\rfloor+1$, then $n$ satisfies: $2^{b-1}\leq n < 2^b$. Therefore,
% $2\leq n/2^{b-2}< 4$.
% \vspace{0.3cm}

% Compute $V_{2}$ and $V_{3}$ using Proposition \ref{Sequence::1}\;
% \For{$s=b-2,\ldots, 1$}
% {	Compute the codes
%     $\begin{array}{ccc}
% 	V_{\left\lfloor ({t+g})/{2^{s-1}}\right\rfloor} & \hbox{ and }&
% 	V_{\left\lfloor ({t+g+1})/{2^{s-1}}\right\rfloor}\end{array}$
%      from the codes
% 	$V_{\left\lfloor({t+g})/{2^s}\right\rfloor}$ and
%     $V_{\left\lfloor({t+g+1})/{2^s}\right\rfloor}$ by applying twice Proposition \ref{Sequence::2}.
% }
% \vspace{0.3cm}
% \underline{Algorithm complexity:} We solve
% $2\lceil\log_2 (t+g)\rceil +2$ systems of linear equations of type (\ref{Problem::1}) and (\ref{Problem::2}).

% \caption{Let $\frac{n}{2}-2>\deg(E)\geq \frac{t+5g+3}{2}$}
% \label{Algorithm::1}
% \end{algorithm2e}
% %%%%%%%%% END algorithm 2

\paragraph{Experimental results}
Our attack has been implemented with \textsc{Magma} \cite{magma}, we summarize in the following tables the average running times for several examples of codes, obtained with an Intel $ \circledR$ CoreTM 2 Duo $2.8$ GHz.
The table includes for each code its base field size $q$, its length $n$,
its dimension $k$, the correction capability $t$ when using error correcting pairs and the key size $\left\lceil (n-k) k \log_2 q\right\rceil \cdot 10^{-3}$ kbits. The last column indicates the running time for the computation
of an ECP for the public code.
 Moreover, the work factor $\mathbf{w}$ of and ISD attack is given.
These work factors have been computed thanks to Christiane Peter's Software
\cite{peters:2010}.

\begin{example}
The {\em Hermitian curve} $\mathcal H_r$ over $\mathbb F_q$ with $q=r^2$ is defined by the affine equation $Y^{r}+ Y = X^{r+1}$.
This curve has $P_{\infty}=(0:1:0)$ as the only point at infinity.
Take $E=mP_{\infty}$ and let $\mathcal P$ be the $n=q\sqrt{q}=r^3$ affine $\mathbb F_q$-rational points of the curve.
Table \ref{Table::1} considers different codes of type $C_L(\mathcal H_r, \mathcal P, E)^{\perp}$ with $n>m>2g-2$.

\begin{table}[h!]
\centering
\begin{tabular}{|c|c|c|c|c|c|c|c|}
\hline
%m
$q$ & $g$ & $n$ & $k$ & $t$ & $\mathbf w$ & key size& time\\
\hline \hline
% m=170
$7^2$ &  $21$ & $343$ & $193$ & $54$ & $2^{84}$ & $163$ kbits & $74$ s \\
\hline
% m = 360
$9^2$ &  $36$ & $729$ & $404$ & $126$ & $2^{182}$ & $833$ kbits & $21$ min\\
\hline
$11^2$ &  $55$ & $1331$ & $885$ & $168$ & $2^{311}$ & $2730$ kbits & $67$ min\\
\hline
\end{tabular}
\medbreak
\caption{Comparison with Hermitian codes}
\label{Table::1}
\end{table}
\end{example}

\begin{example}
The \emph{Suzuki curves} are curves $\mathcal X$ defined over $\mathbb F_q$ by the following equation
$Y^q - Y = X^{q_0} (X^q-X)$
with $q=2q_0^2\geq 8$ and $q_0=2^r$
This curve has exactly $q^2+1$ rational places and a single place at infinity $P_{\infty}$. Let $E=mP_{\infty}$ and $\mathcal P$ be the $q^2$ rational points of the curve.
Table \ref{Table::2} considers a code of type $\C{X}{P}{E}^{\perp}$ with $n>m>2g-2$.

\begin{table}[h!]
\centering
\begin{tabular}{|c|c|c|c|c|c|c|c|}
\hline
$q$ & $g$ & $n$ & $k$ & $t$ & $\mathbf w$ &  key size & time \\
\hline \hline
% m=500
$2^5$ &  $124$ & $1024$ & $647$ & $64$ & $2^{110}$ & $1220$ kbits & $30$ min\\
\hline
% m=600
%$2^5$ &  $124$ & $1024$ & $547$ & $114$ & $2^{150}$ & $2^{47}$ & $1304$ Ko & min\\
%\hline
\end{tabular}
\medbreak
\caption{Comparison with Suzuki codes}
\label{Table::2}
\end{table}
\end{example}

\subsection{In case $t\leq \frac{d^*-1}{2}$, i.e. computing an ECA}
\label{ss:attack_ECA}
In this section we describe how to attack the McEliece cryptosystem
based on AG codes when the integer $t$ is smaller than
$\frac{d^*-1}{2}$. 
{ We first }suppose that
$4g { - 1} \leq \deg (E) \leq \frac {n-4} 2$.
{ The lower degree case will be treated further.}
Then, one can compute the sequences ${(U_i)}_i$ and ${(V_i)}_i$ as follows.

% {\color{blue}\textbf{Explanation - Must be deleted in the final version:}

% By Proposition 7 we should have 
% $$\deg(E + \left(-(2g+1) +1\right)P) \geq 2g \Longrightarrow \deg(E) \geq 2g + 2g +1 -1 = 4g$$

% Or equivalently, by Proposition 20 
% $$2g+1 \leq \deg (E) - 2g +1 \Longrightarrow \deg(E) \geq 2g+1 +2g-1 = 4g$$
% }

\begin{description}
  \setlength{\itemsep}{-.2mm}
  \item[{\bf Step 0}] As explained in Section~\ref{ss:which}, compute
    $V_0$ and $V_{-1}$.
  \item[{\bf Step 1}] By applying iteratively
    Proposition~\ref{prop:comput_Vi}(\ref{item:pb1}) on can compute
    $V_{-2}, \ldots, V_{-2g-1}$ from the knowledge of $V_0, V_{-1}$.
    These computations are possible under the above conditions on
    $\deg (E)$.
  \item[{\bf Step 2}] Compute $U_0, \ldots, U_{2g+1}$ using
    Proposition~\ref{prop:Comp_Ui_from_Vi}(\ref{it:Ui_from_Vi}).
  \item[{\bf Step 3}] Compute $U_{2g+2}, \ldots , U_{4g+2}$ using
    Proposition~\ref{prop:Comp_Ui_from_Ui}(\ref{it:Ui_from_Ui_2}), then
    compute the rest of the sequence ${(U_i)}_i$ using
    Proposition~\ref{prop:Comp_Ui_from_Ui}(\ref{it:Ui_from_Ui_1}).
  \item[{\bf Step 4}] Compute the $V_i$'s for $i \geq 0$ using
    Proposition~\ref{prop:Comp_Vi_from_U_iV_j}.
  \item[{\bf Step 5}] Compute the remaining $V_i$'s (for
    $i < { -2g-1}$) using either
    Proposition~\ref{prop:Comp_Ui_from_Vi}(\ref{it:Vi_from_Ui}) or
    Proposition \ref{prop:Comput_Vi_bis}(\ref{it:compVi_bis}).
\end{description}

{ Assume now that
$2g+1 \leq \deg (E) \leq 4g+1$, one can proceed as follows. First, using
%\sout{Theorem~\ref{thm6-(2)}}}
%{\color{blue}
Corollary~\ref{Corollary::7}}
{ , one can compute $\C{X}{P'}{2E}$ as $\C{X}{P'}{E}^{(2)}$.
Next, since $\deg(2E) \geq 4g+1$, using the above described process, one
can compute the whole filtrations
$$
U_i = \C{X}{P'}{iP} \qquad {\rm and} \qquad V'_i = \C{X}{P'}{2E+iP}.
$$
If $E\geq 0$, then, one can compute any $V_i = \C{X}{P'}{E+iP}$ such that
$2\deg (E) +i < n$ as
$$
\C{X}{P'}{E+iP} = \C{X}{P'}{E} \cap \C{X}{P'}{2E+iP}.
$$
If $E \not \geq 0$, one can compute the $V_i$'s for $i \geq 2g$ using
Proposition~\ref{prop:Comp_Vi_from_U_iV_j}. Next, the $V_i$'s for the
other values of $i$ can be computed using iteratively Proposition~\ref{prop:Comp_Ui_from_Vi}(\ref{it:Vi_from_Ui}).

\begin{remark}
  Note that the case $\deg (E) \leq 2g$ is irrelevant. Indeed, 
  since the Goppa designed distance is $\deg (E) + 2 - 2g$, considering
  codes with $\deg (E) \leq 2g$ would mean that the Goppa designed distance
  is $\leq 2$ and hence no error can be corrected.
\end{remark}
}

% \begin{remark}
% \sout{  The condition on the degree of $E$ can be relaxed by shortening the code
%   as explained in Section~\ref{Extending}.
%   It is also possible to compute the whole sequences if 
%   $3g+2 \leq \deg (E) \leq \frac n 2 - g$ by first computing
%   the codes $V_{-g-1}, \ldots, V_0$ using
%   Proposition~\ref{prop:comput_Vi}(\ref{item:pb1}), then compute
%   the codes $V_1, \ldots, V_g$ using
%   Proposition~\ref{prop:comput_Vi}(\ref{item:pb2}).
%   Then the very same process permits to compute the rest of the sequences.}
% \end{remark}

\paragraph{Complexity} 
Since we have $O(n)$ codes to compute using methods described in
Section~\ref{ss:which}, according to Section~\ref{ss:complexity},
the complexity of the construction of the sequences is $O(n^5)$
if we use a deterministic algorithm and $O(n^{4+\varepsilon})$
if we use the probabilistic shortcut.

% \subsection{Theoretical and Probabilistic complexity}
% \label{sub-complexity}

% As we have already mentioned, the costly part of the attack is the
% computation of a complete (or semi-complete) filtration of the code
% $\C{X}{P}{E}$. That is, solving $\lambda$ linear systems whose cost is
% the computation of the square of a code. So, roughly speaking the cost
% of our attack is about $\mathcal O(\lambda n^4)$ operations in
% $\mathbb F_q$.

% However, it is actually possible to reduce the complexity since most
% of the systems that we have to solve have $b$ equations and $a$
% unknowns with $b\gg a$. With high probability the complete system and
% the subsystem which consist of $a+\varepsilon$ equations (chosen at
% random) has the same solution set. This probabilistic shortcut permits
% the computation of the square of a code in $\mathcal O(n^3)$ and
% reduces the cost of the attack to $\mathcal O(\lambda n^3)$. Therefore
% we have an overall complexity of $\mathcal O(n^4)$.

\subsection{Extending the attack}
\label{Extending}

We have been working under the assumption that $\deg(E) \leq \frac{n}{2}-2$.
In the remainder of this section we
will see how % \sout{by shortening arguments}
this condition can be weakened.

{\subsubsection{By dualizing}
A first manner to address the case $\deg (E) > \frac n 2 - 2$ is to consider
the dual code. Since, from Theorem~\ref{Theorem::2},
 $\C{X}{P'}{E}^\bot = \C{X}{P'}{E^{\perp}}$ with $\deg (E^{\perp}) = n - \deg (E) +
2g - 2$. Next suppose that $\deg (E^{\perp}) \leq \frac n 2 - 2$,
then, using the previous results one can compute the whole
filtrations
$$
U_i = \C{X}{P'}{iP}\qquad {\rm and} \qquad W_i = \CC{X}{P'}{E^\bot+iP}
$$
and there remains to notice that
$$
W_i^\bot = \C{X}{P'}{E - iP} = V_{-i},
$$
which permits to compute the filtration ${(V_i)}_i$.

In summary, this dualization approach permits to treat the case $\deg (E)
\geq \frac n 2 + 2g$. There remains to treat the case where
$$
\frac n 2 - 2 < \deg (E) < \frac n 2 + 2g.
$$
This issue is partially addressed in the next section.
}
{
\subsubsection{By shortening}
 Up to now, we explained how to break the system unless: 
 $$
 \frac n 2 - 2 < \deg (E) < \frac n 2 + 2g.
 $$
%  Note that, in this situation either $E$ or $E^\bot$ have degree
%  $\leq \frac n 2 + g - 1.$ According to the previous section, one
%  can either work on the code or its dual, therefore, there remains to treat the
%  case
%  $$
%  \frac n 2 - 2 < \deg (E) < \frac n 2 + g.
%  $$
% We show how to address this situation by shortening as soon as $\deg (E) > 4g.$
 Note that, according to the previous section, one can either
 work on the code or its dual. Therefore, there remains to treat the
 case
 $$
 \frac n 2 - 2 < \deg (E) < \frac n 2 + g.
 $$
}

\begin{notation}
  Consider the code $C = \C{X}{P}{E}$.  Let $I$ be a subset of $\{1,
  \ldots, n\}$ and $P_I$ the divisor $\sum_{j\in I}P_j$. The code
  $\C{X}{P}{E-P_I}$ is denoted by $C(I)$.
\end{notation}

If we delete the zero positions of the code $C(I)$ we obtain a code of
length $n - |I|$ which is nothing but the shortening of $C$ at
$I$.  The interest of shortening is that
\begin{equation}
  \label{eq:shortening}
  |I| \geq 2 \deg (E) - n +4 \ \ \Longrightarrow \ \ \deg(E - P_I) <
  \frac{n - |I|}{2} \cdot
\end{equation}

Hence, if {$\deg(E) \geq \frac n 2 - 2$}, then for a set of indexes $I$
such that $|I|$ is large enough, the shortened code at $I$ satisfies the
degree constraint.

This can be applied for cryptanalysis as follows.
For instance, suppose we know $V_{i-1}$ and $V_i$ and aim to
compute $V_{i+1}$
using Proposition~\ref{prop:comput_Vi}(\ref{item:pb2}) but unfortunately
$\deg (E) + iP \geq { \frac n 2 - 2}$. Then, choose some sets of indexes
$I_1, \ldots, I_s$ satisfying (\ref{eq:shortening})
and compute the codes $V_i(I_1), \ldots, V_i (I_s),$
and $V_{i-1}(I_1), \ldots, V_{i-1}(I_s)$
by Gaussian elimination. Afterwards, using
Proposition~\ref{prop:comput_Vi}(\ref{item:pb2}), deduce from them
the codes $V_{i+1}(I_1), \ldots, V_{i+1}(I_s)$ and
sum them up. The following statement asserts that this sum of codes
equals $V_{i+1}$ provided some mild conditions on the $I_j$'s
are satisfied.

% \begin{lemma}
% \label{Lemma::Enlarging-prev}
% Let $I$ and $J$ be different subsets of $\{1, \ldots, n\}$ and
% let $E$ be a divisor on the curve $\mathcal X$ with disjoint support from $\{P_1, \ldots, P_n\}$.  Then,
% $$L(E_{I}) \cap L(E_{J}) = L(E_{I\cup J})$$
% \end{lemma}

% \begin{proof}
% Obviously, $L(E_{I\cup J})\subseteq L(E_{I}) \cap L(E_{J})$, since $E_{I\cup J}\leq E_{I}$ and $ E_{I\cup J} \leq E_{J}$.

% Conversely, let $f\in L(E_{I}) \cap L(E_{J})$, then $(f)\geq -E_{I}$ and $(f)\geq -E_J$.
% That is, $\mathrm v_P(f) = - a_P$ where $a_P$ is the coefficient of $P$ in $E$ and $\mathrm v_{P_l}(f) = 1$ for all $l\in I\cup J$. Thus, $(f)\geq -E+\sum_{I\cup J}P_j = -E_{I\cup J}$, which completes the proof.
% \end{proof}

\begin{lemma}
\label{Lemma::Enlarging}
Let $F$ be a divisor of degree less than $n$.
Let $I_1, \ldots, I_s$ be subsets of $\{1, \ldots, n\}$ such that
$\deg (F)  -|\bigcup_{j=1}^s I_j| > 2g-2$.
Then, $$C(I_1 \cap \cdots \cap I_s) = C(I_1) + \cdots + C(I_s).$$
In particular, if $I_1 \cap \cdots \cap I_s = \emptyset$, then
$C = C(I_1) + \cdots + C(I_s).$
\end{lemma}

\begin{proof}
%Set $J \eqdef I_1 \cap \cdots \cap I_s$.
Since $\deg (F) < n$, the evaluation map is injective on $L(F)$.
Hence it is sufficient to prove that 
$$
L(F - P_{I_1 \cap \cdots \cap I_s}) = L(F-P_{I_1}) + \cdots + L(F - P_{I_s}).
$$
We give the proof for the case $s=2$. The general case deduces from that one
using a descending induction on $s$.
First, since $F - P_{I_1}\leq F - P_{I_1 \cap I_2}$ and $F - P_{I_2} \leq F - P_{I_1 \cap I_2}$ we get
$$
L(F-P_{I_1}) + L(F-P_{I_1}) \subseteq L(F-P_{{I_1 \cap I_2}}).$$
Conversely, we will prove that both sides have the same dimension.
For this sake one first observes that 
$$
L(F - P_{I_1}) \cap L(F - P_{I_2}) = L(F - P_{I_1 \cup I_2}),
$$
which entails that
\begin{align*}
\dim (L(F-P_{I_1}) + L(F & -P_{I_2})) = \\
& \dim (L(F-P_{I_1}) ) + \dim  (L(F-P_{I_2})) - \dim (L(F - P_{I_1 \cup I_2})).
\end{align*}
By assumption on $|I_1 \cup I_2|$, all the degrees of the above involved
divisors exceed $2g-2$ and hence, from Riemann Roch theorem,
$$\dim \left(L(F-P_{I_1}) + L(F-P_{I_2})\right)
 =  \deg(F) - g +1 - \left( |I_1| + |I_2| - |I_1 \cup I_2| \right),
$$
which is nothing but $\dim L(F - P_{I_1 \cap I_2})$.
This concludes the proof.
\end{proof}

{
If we go back to the situation
$$
\frac n 2 - 2 < \deg (E) < \frac n 2 + g.
$$
We have $\deg (E) = \frac n 2 - 2 + \varepsilon$ for some
$1 \leq \varepsilon \leq g-1$. Therefore, we need to shorten the code
at least at $2 \varepsilon$ positions. Moreover, for the attack to
work on the shortened code, we need $\deg (E) - 2\varepsilon > 2g$.
In the worst case, $\varepsilon = g-1$ and the attack on the shortened code
is proved to be efficient only if $\deg (E) > 4g - 2$.
}
{
\subsubsection{Are there codes out of the reach of the attack?}
For the codes such that 
$$
\frac n 2 - 2 < \deg (E) < \frac n 2 + g
$$
and such that $\deg (E) \leq 4g-2$, the previously described attack is
not proved to be efficient.  Such codes can be obtained from Garcia
Stichtenoth towers by taking a curve with a large genus $g$ and whose
number of points is $\approx c\cdot g$ for some positive constant $c$.
In this context, one can construct a code of length $n = c'\cdot g$
for some positive constant $c'<c$ and consider a divisor $E$ of degree
$\frac n 2 + g$.  If $c' < 6$ any of the previously described method
cannot be proved to work.

However, for such codes, we can still try to apply the algorithms even
if we have no proof they will provide the expected result.  For a
divisor $A$ of degree $<2g+1$, the code $\C{X}{P}{A}^{(2)}$ may be non
equal to the code $\C{X}{P}{2A}$ but is at least a subcode of it. This
subcode can be used to try to compute subcodes of the form
$\C{X}{P'}{A-iP}$. Moreover, the previous approach using
shortening leaves us many lattitude, since there is a large family of
subsets we can use. Thus, by trying many tuples of
subsets $(I_1,\ldots , I_s)$ we will probably be able to collect vectors of a target space $U_i$ or $V_i$ and after sufficiently many trials, get the whole
target space.
In addition that if this approach turned out to fail on both $\C{X}{P'}{E}$
and $\C{X}{P'}{E^\bot}$, it is always possible to choose another point $P$
in the support $\mathcal P$ and re-try with another $\mathcal P'$.

Despite the lack of proofs in this situation, the existence of a code
for which, the attack would fail for both $\C{X}{P'}{E}$
and $\C{X}{P'}{E^\bot}$, for any choice of shortening and for any choice of
point $P \in \mathcal P$ seems highly improbable.}

%%% Local Variables: 
%%% mode: latex
%%% TeX-master: "Long-Version"
%%% End: 

\section{Subcodes of AG codes}
\label{Section7}
In this section we give a polynomial time attack on the McEliece
public key cryptosystem based on subcodes of AG codes.  Now, our
public key is a non structured generator matrix $\mathbf G$ of a subcode $C$
of $\C{X}{P}{E}^{\perp}$ of dimension $\ell$, together with the error
correcting capacity $t$.  Our attack consists in recovering the
code $\C{X}{P}{E}^{\perp}$ from the knowledge of $C$ and then use one of the
attacks presented in Section~\ref{Section6}. That is, once
$\C{X}{P}{E}^{\perp}$ is recovered, we design an
efficient decoding algorithm for $\C{X}{P}{E}^\perp$ which corrects up to
$t$ errors. This yields a decoding algorithm for our public code $C$.

The genus zero case (i.e. the case of GRS codes) proposed in
\cite{berger:2005} was broken by Wieschebrink \cite{wieschebrink:2010}
as follows:
\begin{itemize}
\item the public key $C$ is contained in some secret
$\mathrm{GRS}_k(\mathbf a, \mathbf b)$;
\item compute $C^{(2)}$ which is, with a high probability,
equal to $\mathrm{GRS}_k(\mathbf a, \mathbf b)^{(2)}$,
which is itself equal to $\mathrm{GRS}_{2k-1}(\mathbf a, \bb * \bb)$.
\item Apply Sidelnikov Shestakov attack \cite{sidelnikov:1992}
 to recover
$\ab$ and $\bb * \bb$, then recover $\bb$.
\end{itemize}
Compared to Wieschebrink's approach, our difficulty is that our attack is not a key-recovery attack but a blind construction of a decoding algorithm. For this reason, even if $C^{(2)}$ provides probably the code $\C{X}{P}{E}^{(2)}$, it is insufficient for our purpose: we need to find $\C{X}{P}{E}$.
This is the reason why we introduced the notion of ${ s}$--closures in Section~\ref{sect:t-closure}.

\subsection{Principle of the attack}
In this section the public key consists in an $\ell$-dimensional
subcode $C$ of the AG code $\C{X}{P}{E}^{\bot}$.
Let $k \eqdef \dim \C{X}{P}{E}$.
We assume that
$$\begin{array}{ccc}
2g+1 \leq \deg(E) \leq \frac{n-1}{2} & \hbox{ and }&
2k-1+g \leq \binom{l+1}{2}
\end{array}.$$
Moreover, assume Conjecture~\ref{conj:squares} to be true.  Then, with
a high probability, we may assume that $C^{(2)} = \C{X}{P}{2E}$ and
hence $\overline{C}^2 = \C{X}{P}{E}$ by Corollary \ref{Corollary::15}.
Therefore, to break the scheme we can proceed as follows.
\begin{description}
\item \textbf{Step 1.} Compute $\overline{C}^2$
by applying Proposition~\ref{Proposition::8-9},
which boils down to Gaussian elimination.
\item \textbf{Step 2.} Apply the polynomial time attack presented in Section~\ref{Section6} to obtain an ECP or an ECA for $\C{X}{P}{E}^{\bot}$.
Which yields a decoding algorithm for $C$.
\end{description}

\begin{remark}
  In case $\deg(E)> \frac{n-1}{2}$,
then the attack can be applied to several shortenings of $C$
whose $2$--closures are computed separately and are then summed up to provide
$\C{X}{P}{E}$. This method is described in Section~\ref{Extending}.
\end{remark}

\paragraph{Complexity:}
The computation of a closure costs $O(n^4)$
operations in $\Fq$. It can be reduced to $O(n^{3+\varepsilon})$ operations
using the probabilistic shortcut presented in Section~\ref{ss:complexity}. 
According to the previous results, the
rest of the attack is at most in $O(n^5)$.

\paragraph{Experimental results}
This attack has been implemented with MAGMA. To this end $L$ random
subcodes of dimension $\ell$ from Hermitian codes of parameters
$[n,k]_q$ were created. It turned out that for all created subcodes a
$t$-ECP could be reconstructed. {\tt Time} represents the average time
of the computation of an Error correcting pair for the public
code obtained with an Intel $ \circledR$ CoreTM 2 Duo $2.8$
GHz. 
 The work factor $\mathbf{w}$ of an ISD attack is given.  These
work factors have been computed thanks to Christiane Peter's Software
\cite{peters:2010}. 

\begin{table*}[h!]
\begin{center}
\begin{tabular}{|c|c|c|c|c|c|c|c|c|}
\hline
%%m=170
$q$ & $n$ & $k$ & $t$ & {\tt Time}  & \hbox{key size} & $\mathbf w$ & $l$ & $L$ \\
\hline \hline
\multirow{3}{*}{$7^2$} & \multirow{3}{*}{$343$} & \multirow{3}{*}{$193$} & \multirow{3}{*}{$54$} &
\multirow{3}{*}{$80$ s}
& $83$ kbits & $2^{30}$ & $50$ & $1000$ \\
& & & &  & $137$ kbits & $2^{43}$ & $100$ & $1000$ \\
& & & &  & $163$ kbits & $2^{62}$ & $150$ & $1000$ \\
\hline
\end{tabular}
\vspace{0.3cm} \\
\begin{tabular}{|c|c|c|c|c|c|c|c|c|}
\hline
%%m=360
$q$ & $n$ & $k$ & $t$ & {\tt Time}  & \hbox{key size} & $\mathbf w$ & $l$ & $L$ \\
\hline \hline
\multirow{3}{*}{$9^2$} & \multirow{3}{*}{$729$} & \multirow{3}{*}{$521$} & \multirow{3}{*}{$19$} &
\multirow{3}{*}{$30$ min}
& $216$ ko & $2^{32}$ & $50$ & $500$ \\
& & & &  & $670$ ko  & $2^{121}$   & $200$ & $500$ \\
& & & &  & $835$ ko & $2^{178}$   & $400$ & $500$ \\
\hline
\end{tabular}
\end{center}
\caption{Running times of the attack over Hermitian codes}
\label{Table::3}
\end{table*}

\subsection{Which codes are subject to this attack?}

The subcode $C \subseteq \C{X}{P}{E}$ should satisfy:
\begin{enumerate}[(i)]
\item\label{item:square} ${\dim C +1 \choose 2} \geq \dim \C{X}{P}{2E}$;
\item\label{item:dimC2E} $2g+1 \leq \deg (E) \leq \frac{n-2}{2}$;
\end{enumerate}
The left-hand inequality of (\ref{item:dimC2E}) is
in general satisfied. On the other hand,
as explained above, the right-hand inequality of (\ref{item:dimC2E})
can be relaxed by using a shortening trick.
Constraint (\ref{item:square}) is more central since a subcode
which does not satisfies it will probably behave like a random code and it
can be checked that a random code is in general $2$--closed. Thus,
computing the $2$--closure of such a subcode will not provide any
significant result. On the other hand, for an AG code of dimension $k$,
subcodes which do not satisfy (\ref{item:square}) have dimension
smaller than $\sqrt{2k}$ and choosing such very small subcodes
and decode them as subcodes of $\C{X}{P}{E}$ would represent
a big loss of efficiency. In addition, if these codes have a too low
dimension they can be subject to generic attacks like information set decoding.

\subsection{Subfield subcodes still resist}
Assume $\F_q$ to be non prime and let $\F$ be a proper subfield of $\F_q$
and $C \eqdef  \C{X}{P}{E} \cap \F^n$.
The point is that $C^{(2)} \subseteq \C{X}{P}{E}^{(2)} \cap \F^n$ and the
$2$-closure of $C$ in general differs from $\C{X}{P}{E}$.
For this reason, subfield subcodes resist to this kind of attacks.
Notice that even in genus zero: subfield subcodes of GRS codes
still resist to filtration attacks unless for the cases presented
in \cite{COT14, COT17}.
Moreover, similarly to the case of classical Goppa codes, some of these
codes are known to have a good designed distance, see for instance
\cite{KatsmanTsfasman, Wirtz, Sticht_Subfield} or
\cite{CouvreurCartier} for another construction based on the Cartier operator.
Therefore, these codes provide a good candidate for a secure generalisation 
of the original McEliece scheme based on classical Goppa codes.

%%% Local Variables: 
%%% mode: latex
%%% TeX-master: "Long-Version"
%%% End: 

\section*{Conclusion}
We gave polynomial time algorithms which permit one to reconstruct either
an error correcting pair or an error correcting array of a given code.
After the works of Faure and Minder \cite{faure:2008}
who broke McEliece scheme
based on codes from hyperelliptic curves of low genus, the present article 
gives a general attack proving that McEliece scheme based on
AG codes from any curve of any genus is totally insecure. 
Moreover, we show that the countermeasure consisting in replacing an AG
code by a random low codimensional subcode is insecure too since the parent
code can be recovered by a computation of a $2$--closure.

On the other hand, similarly to the genus $0$ case, subfield subcodes of
AG codes are completely out of the reach of our attack and remain an
interesting candidate for a generalisation of the original McEliece scheme
based on classical Goppa codes.

%%% Local Variables:
%%% mode: latex
%%% TeX-master: "Long-Version"
%%% End:

\bibliographystyle{abbrv}
\bibliography{Long-Version}

\appendix
\section{From degenerate to non-degenerate}
\label{Degenerate}
In Section \S~\ref{ss:attack_ECP}, we explain how to compute
a subcode $\C{X}{P}{E-F}$ of $\C{X}{P}{E}$, however
this code is degenerated since $F = (t+g)P$ and hence the position
corresponding to $P$ is zero for any codeword of $\C{X}{P}{E-F}$.
This is the reason why we cannot directly perform decoding
on $\C{X}{P}{E}^{\bot}$ and should first decode its puncturing at
$P$. 

In what follows, we explain how to compute another code $\C{X}{P}{E-F'}$,
where $F'$ is linearly equivalent to $F$ , i.e. $F' = F+(h)$ for some rational
and has disjoint support with $\DP$.
It should be pointed out that we do not need to compute $h$
but just prove its existence.
In addition to the previous calculations, the computation of a generator
matrix of $\C{X}{P}{E-F'}$ requires the knowledge of the code $V_{-(t+g+1)}$
which can be obtained thanks to
Proposition~\ref{prop:comput_Vi}(\ref{item:pb1}).

\begin{proposition}
\label{Proposition::degenerated}
Let $\Gm$ be a generator matrix of $V_{-(t+g)}$
of the form
$$
\Gm =\left(
\begin{array}{c|c}
  0\! \! & ~\cb_1~ \\
\hline
 (0)\! \! & ~\Gm_1~
\end{array}
\right),
$$
where $\cb_1 \in \F_q^{n-1}$ and $\left(\begin{array}{c|c}0 & \cb_1\end{array}\right)\in V_{-(t+g)}\setminus V_{-(t+g+1)}$
and $\left(\begin{array}{c|c}(0) & \Gm_1\end{array}\right)$ is a generator matrix of $V_{-(t+g+1)}$.
Then, there exists a rational function $h$ on $\X$ such that the matrix
$$
\Gm'\eqdef \left(
\begin{array}{c|c}
  1\! \! & ~\cb_1~ \\
\hline
 (0)\! \! & ~\Gm_1~
\end{array}
\right)
$$ is a generator matrix for $\C{X}{P}{E-(t+g)P-(h)}$.
\end{proposition}

\begin{proof}
For simplicity suppose that $P = P_1$, i.e. $P$ corresponds to
the first column of the generator matrix.
Let $f\in L(E-(t+g)P)\setminus L(E-(t+g+1)P)$
be the function such that
$(0 ~|~ \cb_1) = \left(f(P_1), \ldots, f(P_n)\right)$.
By definition, $ v_{P}(f) = t+g$. From the weak approximation Theorem \cite[Theorem 1.3.1]{stichtenoth:2009}, there exists a rational
function $h \in \F_q(\X)$ such that
\begin{enumerate}[(i)]
\setlength{\itemsep}{-.2mm}
\item $\forall i \geq 2,\ h(P_i) = 1$;
\item $v_{P_1}(h)=-t-g$ and $hf(P_1)=1$.
\end{enumerate}
Such a function $h$ yields the result. 
\end{proof}

%%% Local Variables: 
%%% mode: latex
%%% TeX-master: "Long-Version"
%%% End: 

\end{document}